\documentclass[a4paper,11pt]{article}
\pdfoutput=1 

\usepackage{jcappub} 

\usepackage[T1]{fontenc} 

\usepackage{graphicx}
\usepackage{amssymb}
\usepackage{ulem}
\usepackage{multirow}
\usepackage{color}
\usepackage{rotating}
\usepackage{placeins} 
\usepackage{cleveref}
\usepackage[font=footnotesize]{caption}

\def\be{\begin{equation}}
\def\ee{\end{equation}}

\def\bi{\begin{itemize}}
\def\ei{\end{itemize}}

\def\ben{\begin{enumerate}}
\def\een{\end{enumerate}}

\def\bt{\begin{tabular}}
\def\et{\end{tabular}}

\def\bc{\begin{center}}
\def\ec{\end{center}}

\def\be{\begin{equation}}
\def\ee{\end{equation}}
\newcommand{\bs}{\boldsymbol}
\newcommand{\bes}{\begin{subequations}}
\newcommand{\ees}{\end{subequations}}
\def\bea{\begin{eqnarray}}
\def\eea{\end{eqnarray}}

\newcommand\mpl{}
\def\tcr{\textcolor{red}}

\def\tcg{\textcolor{green}}

\newcount\hh
\newcount\mm
\mm=\time
\hh=\time
\divide\hh by 60
\divide\mm by 60
\multiply\mm by 60
\mm=-\mm
\advance\mm by \time
\def\hhmm{\number\hh:\ifnum\mm<10{}0\fi\number\mm}

\title{Sensitivity of inflationary predictions to pre-inflationary phases}

\author{Sina Bahrami and \'Eanna \'E. Flanagan}
\affiliation{Department of Physics, Cornell University, Ithaca, NY 14853, USA.}

\emailAdd{sb933@cornell.edu}
\emailAdd{eef3@cornell.edu}

\abstract{How sensitive are the predictions of inflation to pre-inflationary conditions when the number of efolds of inflation is not too large?
In an attempt to address this question, we consider a simple model where the inflationary era is preceded by
an era dominated by a radiation fluid, which is coupled to the inflaton only gravitationally
and which extends back to the Planck era.
We show that there is a natural
generalized Bunch-Davies vacuum state for perturbations to the coupled inflaton-gravity-fluid system at early times.
With this choice of initial state the model predicts interesting deviations from the standard power spectrum of single field slow-roll inflation at large scales.
However, the deviations are too small to be observable in near future CMB observations.
}

\begin{document}
\maketitle
\flushbottom

\def\tcr{\textcolor{red}}
\def\tcb{\textcolor{blue}}
\definecolor{darkgreen}{rgb}{0.0, 0.6, 0.0}
\def\tcg{\textcolor{darkgreen}}

\section{Introduction and summary}\label{intro}




\subsection{Background and motivation}\label{context}

 Inflationary cosmology as we know it today was formulated in the  1980's \cite{guth,steinhardt,linde,star1}. This theory was designed to explain some of the key features of the Universe on large scales while reducing the amount of fine tuning required in initial conditions. In particular it provides a natural explanation for the observed smallness of the Universe's spatial curvature, the observed homogeneity and isotropy on large scales, and the non-observation of magnetic monopoles \cite{weinberg, dodelson, baumann, mukhanov}.

The standard picture of  inflationary cosmology, better known as  single field slow-roll inflation, consists of a single scalar field  (the inflaton) endowed with a canonical kinetic term and a slowly varying potential. The power spectrum of scalar perturbations predicted by this theory is largely consistent with the measured anisotropies of the cosmic microwave background radiation (CMB) \cite{wmap1, planck1, cmbpol}. The measured temperature and polarization anisotropies are consistent with perturbations that are adiabatic, nearly Gaussian and nearly scale invariant, in excellent agreement with the predictions of the standard picture of inflation; see, e.g.\ Ref.\ \cite{ mcallister} for a recent review.
Additionally, the standard picture predicts a
stochastic background of gravitational waves  \cite{star2}. Primordial gravitational
waves leave their imprint on CMB photons by inducing
B-mode polarization patterns.
Future detections of this effect could determine the energy scale of inflation.

How robust  the standard inflationary predictions are to
effects of pre-inflationary phases, if any?
If the currently observable modes exited the horizon many
e-folds after the start of single field slow-roll inflation, then it is
unlikely that any pre-inflationary phases can influence the
dynamics of these modes.  On the other hand, if the horizon exit of
these modes occurred instead just after the start of inflation, is is
conceivable that observational signatures of the pre-inflationary phase
could be encoded in the longest lengthscale modes, those that exited
the horizon first.

There is a well-known discrepancy between
the predicted and observed spectrum of CMB temperature anisotropies
for multipoles $l \lesssim 40$, which is small but
statistically significant in the recent Planck data.
Although speculative, it is conceivable that this discrepancy is due
to evolutionary phases that preceded the single-field, slow roll phases.
Several different scenarios of this type have already been explored in
the literature, including
preceding multifield phases \cite{schutz},
initial non-slow-roll phases \cite{contaldi},
and quantum mechanical
tunneling from other vacua in a string theory landscape
\cite{susskind,bousso}.
Some of the effects observed in these scenarios were also found
in single field models where perturbations are initially
in  non-vacuum states \cite{kk}.

In this paper we address the question of the robustness of the
standard inflationary predictions using a simple yet somewhat speculative
toy model. We assume that the standard inflationary phase is preceded by
an era of power law expansion driven by a perfect
isentropic fluid, whose equation of state is that of radiation
\footnote{This cosmological model has a long history. See for instance
\cite{star3,vilenkin1,vilenkin2} for the study of the effects of this cosmological phase
transition on the stability of quantum fields.}.
The fluid and inflaton are coupled only gravitationally. In
Sec.\ \ref{introduce} we discuss the properties of the background
dynamics of this model, highlighting the existence of two
distinguished transitions in the background evolution. Next, in
Sec. \ref{perturb} we derive the equations of motion for the scalar
and tensor perturbations. In Sec. \ref{properties} we discuss the
adiabaticity of the perturbations in the distant past. We then argue
for the suppression of the entropic perturbation modes in
Sec. \ref{iso}.  We conclude in Sec. \ref{power} and \ref{tensorpower}
with a few numerical case studies of the spectrum of tensor and scalar
perturbations in this model.

\subsection{Summary of results}\label{results}
\begin{itemize}

\item There are two important transitions in this model.  The first occurs
  when the energy density of the fluid coincides with the energy density of
  the inflaton, and signals a change from power law expansion to
  quasi-exponential expansion.  The second, subsequent, transition occurs when the
  breaking of the deSitter time-translation symmetry (parameterized
  by the time derivative ${\dot H}$ of the Hubble parameter) changes from
  being predominantly due to the fluid to being predominantly due to
  the inflaton.  After the second transition the dynamics of the
  perturbation modes becomes indistinguishable from that of standard inflation.

\item Using the analytic background solutions found in
  Sec. \ref{backgroundanalytic}, we show in Sec. \ref{properties} that
  for sufficiently smooth inflationary potentials and ordinary  fluids
  ($\rho+3p >0$),  the fluid and inflaton perturbations decouple at
  early times.  We also  show that  if the fluid has an equation of
  state parameter $w=1/3$ (radiation), all the perturbation modes of
  interest are in an adiabatic regime at early times, and that
  therefore one can define a generalized notion of the Bunch-Davies
  vacuum for the perturbations.

\item We show in Sec. \ref{properties} that the fluid and  inflaton perturbation modes of interest decouple after
the second transition.  Therefore, both the background cosmology and the dynamics of the perturbations become identical to single field slow-roll inflation after the second transition.

\item Our model is incapable of generating significant entropic perturbations, an intuitive fact which we establish in Sec. \ref{iso}. Thus effectively all perturbations are adiabatic.

\item We compute the power spectra of scalar and tensor perturbations
  in the model, and find interesting deviations from standard
  predictions at long lengthscales.  In particular there are
  oscillatory features in the low $l$ end of both spectra
  (similar to those seen in trans-Planckian models), whose
  amplitude depends on the ratio of the fluid energy density to the
  inflaton energy density at the epoch when the $l=2$ mode left the
  horizon.  In addition the scalar and tensor spectra are suppressed
  at low multipoles.

\item For some values of our model parameters, the observed modes do
  not have a unique adiabatic vacuum at early times, because the mode
  evolution has not yet become adiabatic when followed backwards in time
  before the Planck scale is reached.  In this portion of parameter
  space, our model loses predictability.  By restricting the values
  of the model parameters, we can avoid this portion of parameter
  space.  However, when we do so, the deviations from standard
  predictions are constrained to be small, of order a few percent.

\item We have not performed an analysis of the detectability of our
  predicted effects.  However, rough estimates suggest that in the
  regime where our model is predictive, the effects are too small to
  be seen in near-future CMB experiments.

\end{itemize}

\section{The fluid-inflaton model} \label{introduce}

We consider a cosmological model for before and during inflation
consisting of a perfect isentropic
fluid and a scalar inflaton field $\varphi$ that are coupled  only
gravitationally.
The  action for the model in the metric signature $(-+++)$ is
\begin{equation} \label{act00}
S=  \int  \sqrt{-g}\: d^4 x \: \Big[  \frac{1}{2 } R+  \mathcal{L}_{f}
- \frac{1}{2} (\nabla \varphi)^2  - V(\varphi)
\:\Big],
\end{equation}
where $R$ is the Ricci scalar, $\mathcal{L}_f$ is the fluid Lagrangian,
$V(\varphi)$ is the inflaton potential, and we are using units with
 $\hbar = c = 1 = 8 \pi G$.  The corresponding equations of motion are
the Klein-Gordon equation
\begin{equation} \label{phi1}
\nabla_{a} \nabla^{a} \varphi + V_{,\varphi}(\varphi) = 0,
\end{equation}
the Einstein equation
\begin{equation}
 G_{ab}=  T^{\text{scalar}}_{ab} + T^{\text{fluid}}_{ab},
\end{equation}
where
\begin{equation}
T^{\text{scalar}}_{ab}  = \nabla_{a} \varphi \nabla_{b} \varphi  - \frac{1}{2} g_{ab} \big[ (\nabla \varphi)^2+2 V(\varphi) \big]
\end{equation}
and
\begin{equation}
T^{\text{fluid}} _{ab} = (\rho_f + p_f ) u_{a} u_{b} + p_f g_{ab}
\end{equation}
are the inflaton and fluid stress-energy tensors,
and the fluid equation
\be
\nabla_a T^{\text{fluid}\, ab} =0.
\ee
Here $\rho_f$, $p_f$ and $u^a$ are the fluid density, pressure and
4-velocity.  We will later specialize the equation of state $p_f =
p_f(\rho_f)$ of the fluid
to be that of radiation, $p_f = \rho_f/3$.

In this section we will analyze the background cosmological solutions
of this model,
and in Sec.\ \ref{perturb} below we will analyze the behavior of perturbations.

\subsection{Background cosmology}

For a homogeneous, isotropic and spatially flat background spacetime
we write the metric as
\begin{equation}
ds^2= - a(\eta)^2 [ d\eta^2 - (d \mathbf{x})^2 ],
\end{equation}
where $a(\eta)$ is the scale factor and $\eta$ is the conformal time.
We write the background solutions for the inflaton and the fluid as $\varphi (\eta)$ and $ \rho_f ^0 (\eta)$, with the background fluid pressure being $p^0 _f = p_f (\rho_f ^0)$. These solutions obey the two Friedmann equations
\begin{equation}\label{fried1}
3 \frac{\mathcal{H}^2}{a^2}= \rho^0 _f + \frac{1}{2 a^2} \varphi^{\prime2} + V(\varphi)
\end{equation}
\begin{equation}\label{fried2}
\frac{1}{ a^2} \big(2 \mathcal{H}^{\prime}+ \mathcal{H}^2 \big) = - p^0 _f - \frac{1}{2 a^2} \varphi^{\prime2} + V(\varphi),
\end{equation}
where primes denotes differentiation with respect to $\eta$ and
$\mathcal{H} \equiv a'/a $
is the conformal Hubble
parameter, and the Klein-Gordon and fluid equations
\begin{equation}\label{klein}
\varphi^{\prime\prime}+ 2 \mathcal{H} \varphi^{\prime} + a^2 V_{,\varphi}(\varphi)=0,
\end{equation}
\begin{equation}\label{cont}
\rho^{\prime 0} _f + 3 \mathcal{H} (\rho^{0}_f + p^0_f ) =0.
\end{equation}
The continuity equation \eqref{cont} yields
\begin{equation}\label{8}
\rho^{0}_f = \frac{\mathcal{E}_0}{a^{3(1+w)}},
\end{equation}
for an equation of state $p_f = w \rho_f$, and a constant of integration $\mathcal{E}_0$. In our analysis of the background dynamics we will mainly use the first Friedmann equation \eqref{fried1} and the two equations \eqref{klein} and \eqref{cont}.

\subsection{The background dynamics and parameters} \label{timescale}

The background fluid-inflaton Lagrangian depends on a number of
parameters, namely the equation of state parameter $w$ of the fluid and all
the coupling parameters in the inflaton potential.  In addition, the space
of solutions of the differential equations
\eqref{fried1}, \eqref{klein} and \eqref{cont} is four dimensional,
requiring four initial conditions to determine a solution.
One of these parameters is gauge, since we can resale the coordinates
to set the value of $a$ at the initial time to unity.
The remaining three parameters can be taken to be
the initial value and initial time derivative of the inflaton,
and the initial ratio $\rho_{\varphi}/\rho_{f}$ of the energy densities of the inflaton and fluid.

In exploring this space of model parameters, our strategy will be to
choose models that agree with observations in the absence of a fluid,
and then gradually modify the model by dialing the parameters related
to the inflaton at some initial time when the energy density of the
fluid is Planckian. In particular we will restrict attention to models
in which, in the absence of a fluid, inflation occurs at the GUT scale
of $\sim 10^{16} \, {\rm  GeV}$.
We will use as an example the large field mode $V = m^2 \varphi^2/2$
to
construct numerical examples of the power spectra of the scalar and
tensor perturbations in sections \ref{numericscalar} and
\ref{numerictensor}.

\begin{figure}[h]
\centering
\includegraphics[width= 1\textwidth]{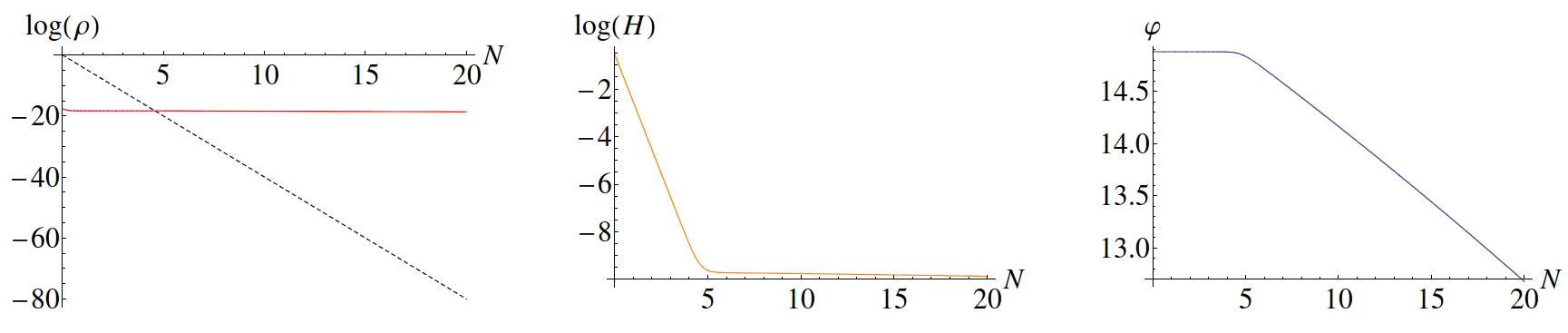}
\caption{{\it Left:} Logarithm of energy density $\rho$ as a function of
  number $N$ of efolds of inflation.  The radiation fluid is shown in
  black and the inflaton in red.  Note that the radiation fluid is
  initially dominant but rapidly becomes subdominant.  {\it Middle:} Log of the Hubble parameter
  $H$ as a function $N$.  Once the inflaton starts
  to dominate at $N \sim 5$, the Hubble parameter becomes
  approximately constant and the expansion becomes quasi-exponential.
  {\it Right:} The inflaton field $\varphi$ as a function of $N$.
  Note that the evolution of the inflaton is nearly frozen during the
  fluid dominated era.}
\label{fig1}
\end{figure}

To illustrate the qualitative features of the the background dynamics,
Fig.\ \ref{fig1} shows
the evolution of the fluid and inflaton energy densities, the Hubble
parameter $H = \mathcal{H}/a$, and the inflaton $\varphi$.
For this example we have taken
$V(\varphi)=  m^2 \varphi^2 /2$ with $m  =  10^{-5} $,
and for the fluid we have taken $w = 1/3$ (radiation)
and used a Planckian initial energy density.
The initial inflaton kinetic and potential energies are
\begin{equation}
\frac{\varphi'^2}{2 a^2} \approx V(\varphi) \approx  10^{-8},
\end{equation}
which is about the grand unification energy density (note that as stated previously we are using units with $m_p =1$). As seen in Fig. \ref{fig1}, the background dynamics starts in a radiation dominated era. The energy densities become equal at  $N \approx 5$, after which the energy density of the radiation keeps decreasing rapidly while the background dynamics becomes similar to that of single field slow-roll inflation. The ensuing near constancy of the Hubble parameter bears evidence to the Universe expanding quasi-exponentially, as expected from the standard picture of inflation.

During the early, fluid dominated era, the inflaton field evolves very
little and is effectively frozen.  This can be seen in the third panel
of Fig. \ref{fig1}.  The freezing of the inflaton is easy to
understand:
defining $N = \ln a$, we have $d \ln \varphi / d N = {\dot \varphi} /
(\varphi H)$.  When $\rho_f \gg \rho_\varphi$, the fluid dominates the
right hand side of the Friedman equation (\ref{fried1}), giving
\be
\frac{d \ln \varphi}{d N} \sim \frac{\mpl {\dot \varphi} }{\varphi \sqrt{\rho_f}}
\lesssim \mpl \frac{1}{\varphi} \sqrt{\frac{\rho_\varphi}{\rho_f}}.
\ee
Thus we have $d \ln \varphi/dN \ll 1$ whenever $\rho_\varphi \ll
\rho_f$, as long as the inflaton field is larger than the
Planck scale.  One can also think of this effect in terms of an
enhanced Hubble damping: the Hubble damping term in the Klein Gordon
equation (\ref{klein}) is much larger than it would be without the
fluid, since  $H^2$ is much larger than
$\rho_\varphi$.
Because of this enhanced damping, the inflaton kinetic energy rapidly
becomes much smaller than its potential energy.
For example, in the model shown in Fig.\ \ref{fig1}, the kinetic and
potential energies are initially equal, but by the time of
fluid-inflaton equality the kinetic energy has decreased by more than
three orders of magnitude.  This then implies that the inflaton
dominated phase starts out in the slow-roll regime: the value
of the slow roll parameter $\varepsilon_{\varphi} \equiv  \varphi'^2 /(2
\mathcal{H}^2)$ when the inflaton begins to dominate is $\sim 10^{-3}$.

\begin{figure}[h]
\centering
\includegraphics[width=0.7\textwidth]{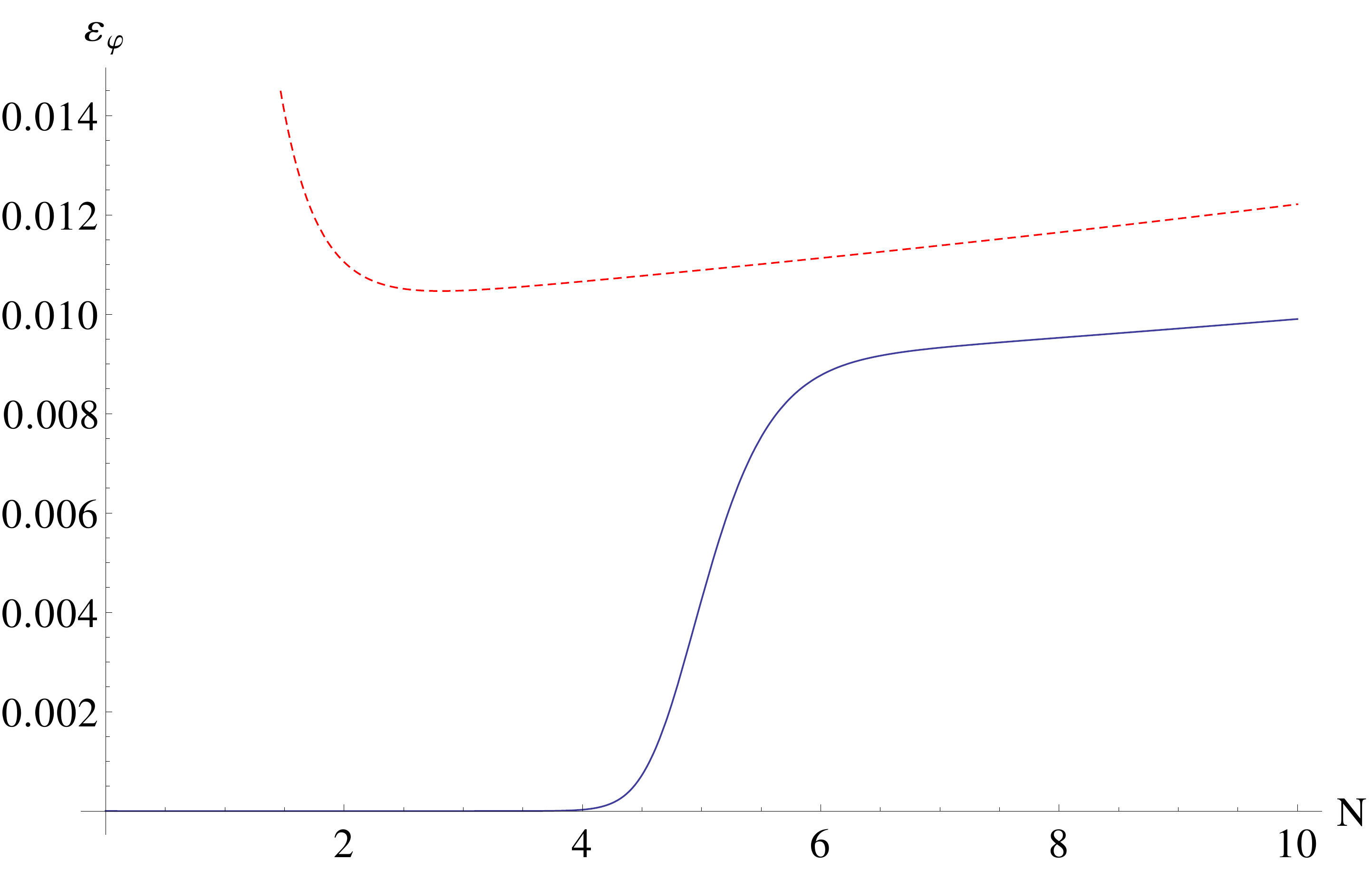}
\caption{The inflaton slow roll parameter $\varepsilon_{\varphi}$
as a function of number $N$ of efolds, for the fluid-inflaton model
with potential $V(\varphi) = m^2 \varphi^2/2$ (blue), and for the
corresponding inflaton-only mode (dashed red).
 In this example, the
initial fluid energy density is the Planck scale, while that of the
inflaton is about the grand unification scale $10^{-2}$.
The slow roll parameter in the inflation-only model starts at some
value set by the initial conditions, and rapidly decays to the
asymptotic track determined by the potential.  By contrast, in the
fluid-inflation model, the slow roll parameter is initially driven to
a very small value during the fluid domination regime, and then
relaxes upwards to the asymptotic track determined by the potential in
the inflaton dominated regime.}
\end{figure}

As discussed in the introduction, there are two important transitions that occur in the fluid-inflaton
model.  The first is the transition already discussed of $\rho_\varphi
\sim \rho_f$, where the background expansion changes from the fluid
dominated power law expansion to the inflaton dominated
quasi-exponential expansion.  After this first transition there is an
approximate de-Sitter expansion.

\begin{figure}[h]
\caption{The comoving scale $a H$ versus number $N$ of efolds for a
  fluid with an equation of state parameter $w=1/3$ and an inflaton
  with a $\varphi^2$ potential. The two important transitions are
  shown. }
\centering
\includegraphics[width=0.9 \textwidth]{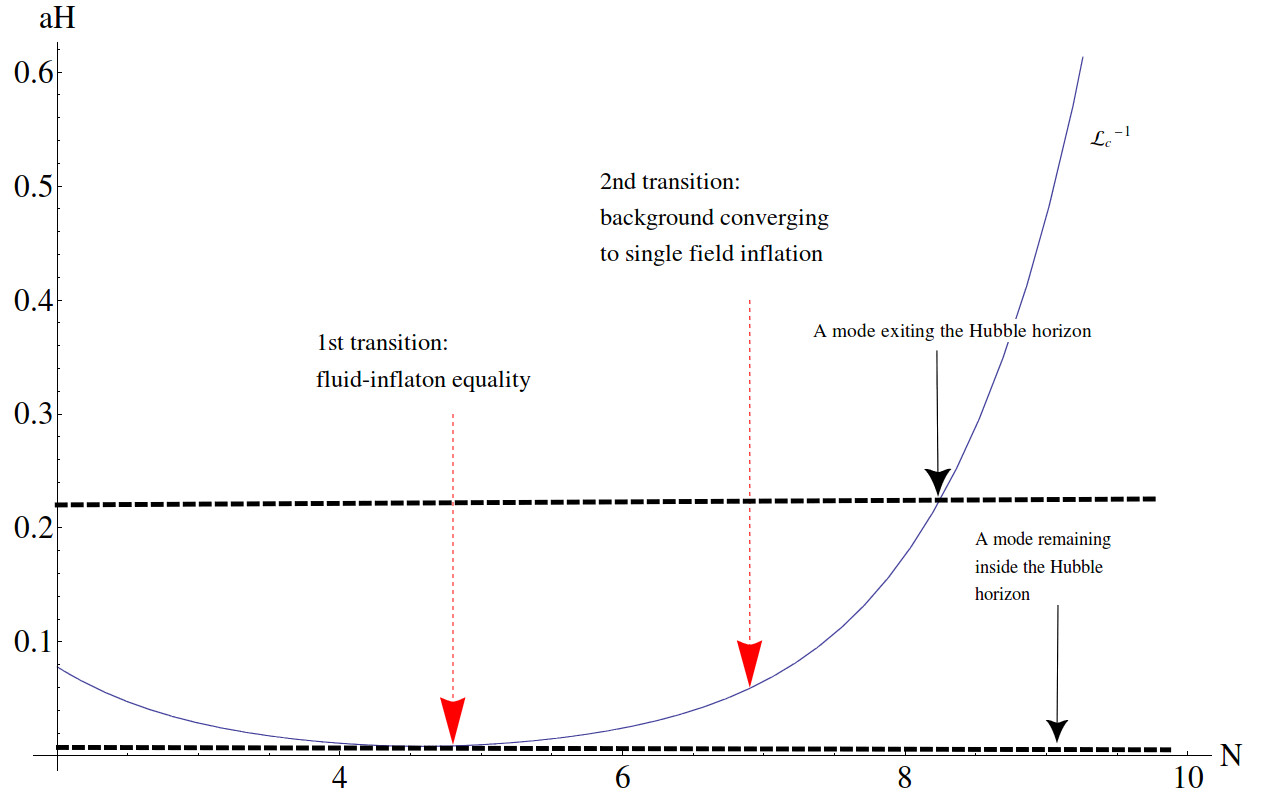}
\end{figure}

 At the first transition, the comoving horizon scale $\mathcal{L}_c
 \equiv  \mathcal{H}^{-1}$ has a local maximum. It increases during the fluid dominated era, since
\begin{equation}
\frac{d}{da} \mathcal{L}_c \propto \rho +  3 p,
\end{equation}
and during this era $\rho$ and $p$ are dominated by the fluid which we
assume obeys the strong energy condition $w>-1/3$. It decreases during
the inflaton dominated era as usual.  Because of the existence of a
local maximum, there exists a minimum comoving wavenumber
$k_{\text{min}}$ such that modes with $k \geq k_{\text{min}}$ leave
the Hubble horizon during the inflaton dominated phase, but modes with
$k \leq k_{\text{min}}$ are outside of the Hubble horizon throughout
the fluid and inflaton dominated phases.
Using
\begin{equation}
3 \mathcal{H}^2 = \frac{\mathcal{E}_0}{a^{1+3w}}+ \Lambda_0 a^2,
\end{equation}
(where we have approximated $\rho_{\varphi} \approx \Lambda_0$ at early times), minimizing with respect to $a$ and noting that we choose $\mathcal{E}_0 = 1$ at $a=1$ gives
\begin{eqnarray}\label{min}
k_{\text{min}} \sim \Lambda_0 ^{\frac{1+3w}{6(1+w)}}.
\end{eqnarray}
For example, for radiation with $w=1/3$ we have $k_{\text{min}} \sim   \Lambda_0 ^{1/4}$.

The comoving Hubble horizon defined above is completely determined by
the background dynamics. However, in this model there are different
scales analogous to the Hubble horizon that are more relevant in
discussing the evolution of the perturbations. To see this,
we use some of the results derived in Sec.\ \ref{perturb} below.
Let us initially
neglect the interaction terms between the fluid and the inflaton
perturbation modes in Eq. \eqref{mateqm}. We see that each mode obeys
a  harmonic oscillator equation with the effective masses
$m_{\varphi}^2 (k, \eta)$ and $m_{f}^2 (k, \eta)$ given by the
diagonal elements of the matrix $\bs{\Omega}$ defined in
Eq. \eqref{matrix}. One can see that the perturbations evolve
differently in two distinct regimes. The first regime is when
$m_{\varphi} ^2 \sim m_{f} ^2 \sim  k^2$, and the second regime is
when $k^2$ is much smaller than $|m_{\varphi} ^2 -k^2|$ and $|m_f ^2 -
c_s ^2 k^2 |$. The transition occurs roughly when
\be
k^2 \sim |m_{\varphi} ^2-k^2| \ \text{or} \  |m_f ^2/c_s^2 -k^2 |.
\ee
Based on this transition property, we define
\be
\mathcal{L}_{c-\text{inflaton}} \equiv \frac{1}{\sqrt{|k^2 - \Omega_{11}|}}, \hspace{1cm} \text{and} \hspace{1cm} \mathcal{L}_{c-\text{fluid}} \equiv \frac{1}{\sqrt{|k^2 - \Omega_{22}/c_s^2|}}
\ee
in analogy with the Hubble horizon.

\begin{figure}[h]
\centering
\includegraphics[width=0.8\textwidth]{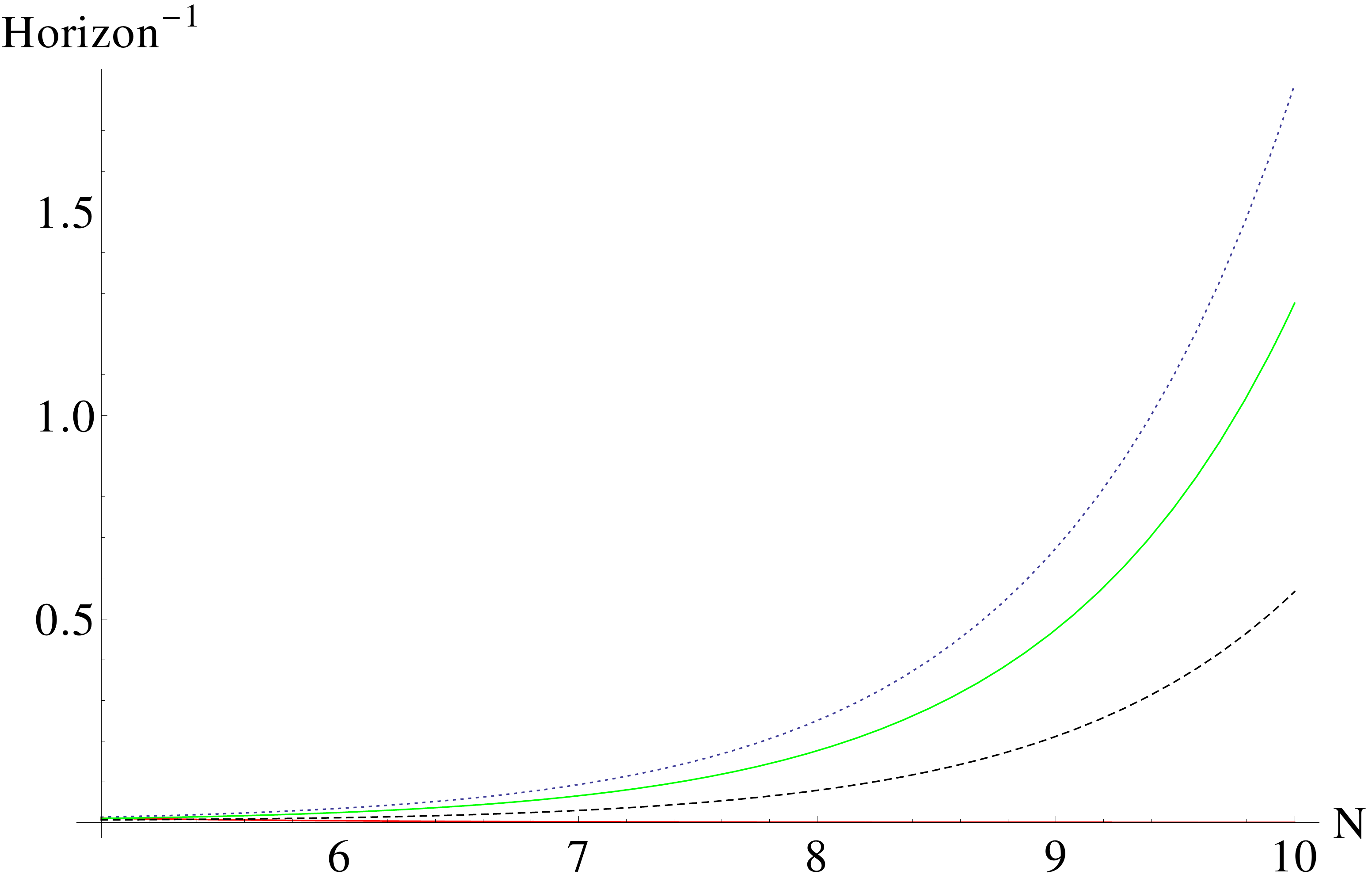}
\caption{The inverses of the three horizon scales $\mathcal{L}_c$, $\mathcal{L}_{c-\text{inflaton}}$ and $\mathcal{L}_{c-\text{fluid}}$ versus number of efolds $N$. The green plot is the plot of $\mathcal{L}^{-1} _c = \mathcal{H}$, which is a background function used to define a local-in-time horizon. The dotted blue plot is the plot of the inverse of the inflaton adiabatic horizon scale $\mathcal{L}^{-1} _{c-\text{inflaton}}$,  and the dashed black plot is the plot of the inverse of the fluid adiabatic horizon scale $\mathcal{L}^{-1}_{c-\text{fluid}}$ associated with a fluid with an equation of state parameter $w=0$. The adiabatic horizon scales roughly signify the boundary between the two distinct regimes for the perturbations; namely the highly oscillatory regime and usually the non-oscillatory regime. We have also included a plot of the inverse of the fluid adiabatic horizon scale for  the case of radiation $w = 1/3$ (the red plot vaguely seen at the bottom).  Note that the radiation case is an exception due to being conformally invariant in any (conformal) power law cosmology. }
\label{horizon}
\end{figure}

Nevertheless, we will indicate in Sec. \ref{vacuum} that the scalar modes with wavenumbers larger than $k_{\text{min}}$ are the ones that are of primary interest to us. These scalar perturbation modes are initially in an adiabatic regime.  Furthermore, it can be shown that after the first transition, while the background geometry is nearly de Sitter, the above defined adiabatic horizon scales are within a factor unity of the Hubble horizon. Thus anytime we refer to a mode exiting the horizon we shall simply mean its wavenumber $k$ becoming equal to $\mathcal{H}$. See Fig. \ref{horizon} for a plot of all three horizon scales.

We now turn to a discussion of the second important transition in the model.
This transition occurs when the breaking of the time translation symmetry (characterizing the deviation of the spacetime from the de Sitter spacetime) switches from being predominantly due to the fluid to predominantly due to the inflaton. Recall that
\begin{equation}\label{broken}
\dot{H} = \frac{1}{a} \Big(\frac{\mathcal{H}}{a} \Big)' = - \frac{1}{2} \Big[ (1+w) \frac{\mathcal{E}_0}{a^{3(1+w)}} + \frac{\varphi'^2}{a^2} \Big],
\end{equation}
where $H$ is the Hubble parameter  and dot denotes differentiation with respect to the coordinate time variable $t$.
Before the transition, the first term in Eq. \eqref{broken} dominates,
and afterwards the second term dominates . One expects the background
dynamics to begin to converge to that of single field inflation once
the time translation symmetry is broken by the inflaton kinetic
energy. We can also quantify this transition in terms of the fluid and
the inflaton contributions $\varepsilon_{f}$ and
$\varepsilon_{\varphi}$ to the primary slow roll parameter
$\varepsilon \equiv  - \dot{H}/H^2$, given by
\begin{equation} \label{sl}
\varepsilon_{f} \sim \frac{ \rho_f}{H^2} , \hspace{1cm} \varepsilon_{\varphi} \sim \frac{\dot{\varphi}^{2}}{H^2}.
\end{equation}
The second transition occurs when $\varepsilon_{\varphi} \gg \varepsilon_{f}$.
To be more precise, we note that slow-roll inflation requires the smallness of another parameter known as the second slow-roll parameter $\eta \equiv - \ddot{H}/(2 \dot{H} H) $. This parameter  ensures the smallness of the inflaton field acceleration.  Like the primary slow-roll parameter $\varepsilon$, this parameter also has contributions due to the fluid, $\eta_{f}$, as well as the inflaton, $\eta_{\varphi}$, that are given by
\begin{equation}
 \eta_{f} \sim \frac{\rho_f}{\varepsilon H^2 }, \hspace{1cm} \eta_{\varphi} \sim \bigg |\frac{\ddot{\varphi} \dot{\varphi}}{\varepsilon H^3 }\bigg |.
\end{equation}
We expect to recover  a single field slow-roll background dynamics by the time $\varepsilon \approx \varepsilon_{\varphi}$ and $\eta \approx \eta_{\varphi}$. Thus, we take these two conditions to be the defining conditions for the second transition. These conditions together place an upper bound on the energy density of the fluid at the moment of the second transition, that is
\begin{equation} \label{rho}
\rho_f \lesssim \varepsilon_{\varphi} \eta_{\varphi}  H^2.
\end{equation}
In what follows we shall take $\rho_f = \varepsilon_{\varphi} \eta_{\varphi}  H^2$ as the defining moment for the second transition. The fluid energy density falls below  the upper bound of Eq. \eqref{rho} at an e-fold  around
\begin{equation}
\log \Bigg[{\Bigg( \frac{1}{\varepsilon_{\varphi} \eta_{\varphi} \Lambda_1}\Bigg)^{\frac{1}{3(1+w)}}}\Bigg],
\end{equation}
where $\Lambda_1$ is the approximate value of the inflaton energy density during inflation.
For a fluid dominated phase with $w = 1/3$  and an $\varepsilon_{\varphi} = | \eta_{\varphi}| = 0.01$, we find that after $N \approx 6.9$ e-folds  the background dynamics is practically identical to that of  a single field slow-roll inflation.

\subsection{Analytic solutions to the background equations in asymptotic limits}\label{backgroundanalytic}
The background  Eqs. \eqref{fried1} to \eqref{klein} can be solved analytically in the limits when either one of the energy densities is sufficiently subdominant to the other one. The strong Hubble damping imposed on the motion of the inflaton field up until the second transition allows one to approximate the energy density of the inflaton as a cosmological constant to leading order.  This means that in either regime, we expect to approximate the inflaton potential using the first few terms in its Taylor series expansion
\begin{equation}\label{potential}
V(\varphi)= \Lambda_* + V'(\varphi_*) (\varphi- \varphi_*) + \mathcal{O}\big[(\varphi-\varphi_*)^2 \big],
\end{equation}
where $\varphi_*$ is the value of the inflaton field either at the beginning of the fluid dominated era, or at the onset of the second transition, when the inflaton energy density becomes dominant. This property will be particularly useful when we discuss  the decoupling and adiabaticity of the perturbation modes at early times in Sec. \ref{properties}.

Based on this insight, we regard the background cosmology at early times to be the fluid plus a cosmological constant and we study the  perturbations around this model induced by the variations in the inflaton field.  Examining Eqs. \eqref{fried1} and \eqref{klein} in this limit reveals that  we can solve the background equations perturbatively using the following expansions for the scale factor and the inflaton field
\be
a = \hat{a} \big[1 + \sigma \delta a ^{(1)} + \sigma ^2 \delta a^{(2)} + \mathcal{O}(\sigma ^3) \big], \hspace{1cm} \varphi = \varphi_0 \big[1  + \sigma  \delta \varphi^{(1)} + \sigma^{2} \delta \varphi ^{(2)} + \mathcal{O}(\sigma^{3}) \big],
\ee
where $\varphi_0$ is the initial value of the inflaton field, $\hat{a}$ is the scale factor due to the dominant fluid, and $\sigma$ is the perturbation  variable taken to be the dimensionless ratio of the energy density of the inflaton over the fluid when the fluid is Planckian. It turns out that the background equations are easily solved using this perturbation scheme. What makes these equations particularly simple to solve is the fact that to each order in $\sigma$, the Eqs. \eqref{fried1} and \eqref{klein} decouple. In other words, the $\sigma ^{n-1}$ order term in $a$ is used to solve for the $\sigma^{n}$ order term in $\varphi$. The reason is simple to see; as we commented earlier, this perturbation method is really a perturbation around the fluid plus cosmological constant model generated by the variations in the inflaton field. Indeed the first order correction to the scale factor, $\delta a ^{(1)}$, is due to the zeroth order inflaton field acting as a cosmological constant with $\rho_{\varphi} = V(\varphi_0) = \Lambda_0$. For fluids with $w > -1/3$, one finds that $\delta a^{(1)}$ decays like $\eta ^{\frac{8+ 6 w}{1+3w}}$ as $\eta$ (or $\hat{a}$) approaches zero.

Using the zeroth order scale factor $\hat{a}$, one then finds that there exists a solution for which $\delta \varphi^{(1)}$ decays like $\eta^{\frac{6+6w}{1+3w}}$. This confirms that the expansion \eqref{potential} is valid at early times. Other solutions for $\delta \varphi^{(1)}$ diverge in the limit of $\eta$ going to zero, spoiling the fluid domination at early times. We therefore specialize to this class of solutions for the inflaton field.

To provide a concrete example, we  apply the above formalism to the case of a radiation fluid ($w= 1/3$) coupled to an inflaton with $V(\varphi) = m^2 \varphi^2 /2$. In this case, we find
\bea \label{pastperturb}
&&  a(\eta) = \sqrt{ \frac{\mathcal{E}_0}{3}} \eta \bigg [ 1+ \frac{\varphi_0 ^2 m^2 \mathcal{E}_0 }{180} \eta^4 + \frac{\varphi_0 ^2 (5 \varphi_0 ^2 -12) m^4 \mathcal{E}_0 ^2 }{194400} \eta ^8 + ... \bigg], \nonumber\\
&& \varphi(\eta) =  \varphi_0 \bigg [ 1 - \frac{1}{60} m^2 \mathcal{E}_0 \eta^4 - \frac{(2 \varphi_0 ^2 -15) m^4 \mathcal{E}_0 ^2}{194400} \eta^8 + . . . \bigg ].
\eea

Next, we analyze the background cosmology after the second transition, when the background dynamics begins to converge to single field slow-roll inflation. As a result of the initially-fluid-induced Hubble damping, the inflaton field behaves as a cosmological constant to a very good approximation for a while even after the second transition.  Thus, we can employ a similar perturbation scheme to study the perturbations to this cosmological constant model induced by both the inflaton field variations and the remaining fluid energy density. To do this, consider the following expansions for the scale factor and the inflaton field,
\begin{equation} \label{expfut}
a  = \bar{a} \big[ 1+ \bar{\varepsilon}_{\varphi}  \Delta a^{(1)}   + \bar{\varepsilon}^2_{\varphi}  \Delta a^{(2)}  + \mathcal{O} (\bar{\varepsilon}^3_{\varphi}) \big],   \hspace{1cm} \varphi= \varphi_1  \big[ 1+ \bar{\varepsilon}_{\varphi}  \Delta \varphi^{(1)}  + \bar{\varepsilon}^2_{\varphi}  \Delta \varphi^{(2)} + \mathcal{O} (\bar{\varepsilon}^3_{\varphi}) \big],
\end{equation}
where $\varphi_1$ is the value of the inflaton field at the second transition, $\bar{a}$ is the scale factor due to the dominant cosmological constant, and the perturbation parameter $\bar{\varepsilon}_{\varphi}$ is the inflaton slow-roll parameter introduced in Eq. \eqref{sl} that is evaluated at the second transition. As mentioned in the last subsection, this parameter breaks the time-translation symmetry during the ensuing slow-roll inflation.

To proceed, we have to determine at what orders in $\bar{\varepsilon}_{\varphi}$ the added fluid will contribute to the perturbative expansions in Eq. \eqref{expfut}. This can be understood using the upper bound derived for $\rho_f$ in Eq. \eqref{rho}. There we found that once $\rho_f \sim \bar{\varepsilon}_{\varphi} ^2 \Lambda_1$, the background dynamics begins to converge to that of single field inflation (we have taken $\eta_{\varphi} = \varepsilon_{\varphi}$ for simplicity). This regime is precisely where our perturbation scheme is valid. It can then be seen from Eqs. \eqref{fried1} and \eqref{klein} that the leading fluid corrections to the scale factor and the scalar field are at most of order  $\bar{\varepsilon}_{\varphi} ^2$.

It actually turns out that the leading  fluid correction to the scale factor is of order  $\bar{\varepsilon}_{\varphi} ^2$, while its leading correction to the scalar field is of order  $\bar{\varepsilon}_{\varphi} ^3$. This is a result of the same decoupling of Eqs. \eqref{fried1} and \eqref{klein} to each order in perturbation that also occurred in the distant past regime. In this case, the decoupling occurs because the zeroth-order inflaton field (the cosmological constant) induces variations in itself via the Klein-Gordon equation \eqref{klein}, i.e. the first order scalar field correction $\Delta \varphi^{(1)}$ is generated by the cosmological constant $V(\varphi_1) = \Lambda_1$. Therefore, it is easy to see that  the $\bar{\varepsilon}_{\varphi} ^{n-1}$ order term in $a$ can be used to solve for the  $\bar{\varepsilon}_{\varphi} ^{n}$ order term in $\varphi$.

For the  case of a radiation fluid ($w= 1/3$) coupled to an inflaton with $V(\varphi) = m^2 \varphi ^2 /2$
a straightforward calculation gives
\bea \label{futureperturb}
&&  a(\eta) = \sqrt{ \frac{3}{\Lambda_1}} \Delta \eta^{-1} \bigg [ 1- \frac{27-21 \Delta \eta - 7 \Delta \eta^4 + \Delta \eta^7 + 42 \Delta \eta \log{|\Delta \eta|}}{42 \Delta \eta} \bar{\varepsilon}_{\varphi} + ... \bigg], \nonumber\\
&& \varphi(\eta) = \varphi_1 \bigg[ 1+ \frac{2-2 \Delta \eta ^3 +6 \log{|\Delta \eta|}}{6} \bar{\varepsilon}_{\varphi}+... \bigg].
\eea
Here $\Lambda_1 = m^2 \varphi_1 ^2 /2$ and $\Delta \eta \equiv \eta - \eta_{\text{end}}$, with $\eta_{\text{end}}$ being the conformal time at which the scale factor $\bar{a} = \sqrt{3}/ (\sqrt{\Lambda_1} \Delta \eta)$ becomes infinite. Also, in the above calculation we have set $\Delta \eta$  equal to one at the second transition.

We will use the equations derived here in Sec. \ref{properties} to show that the radiation  and inflaton scalar perturbations will dynamically decouple in the distant past (radiation dominated) and the future (inflaton dominated) regimes.

\subsection{Post-inflation dynamics: reheating} \label{reheating}

Conventionally, inflation ends when the primary slow-roll parameter $\varepsilon$ becomes equal to unity. At this time, the Universe no longer expands quasi-exponentially since the kinetic energy and the potential fluctuations of the inflaton field are large enough to completely break the de Sitter symmetry. Once this occurs, the Universe enters a phase commonly known as reheating. During this phase, the inflaton field decays to other particles, though the exact mechanism  remains unknown and is largely unconstrained [See \cite{mukhanovcos} for a good review].

The simplest approach to understanding reheating is to assume that inflaton decays to a single component fluid with an equation of state parameter  $\mathcal{W}_{re}$.  The decay occurs while inflaton, now in the form of non-relativistic particles, oscillates at the bottom of its potential. The Klein-Gordon equation \eqref{klein} is then corrected to take into account the decay process,
\begin{equation}\label{decay}
\rho'_{\varphi}+3 \mathcal{H}\rho_{\varphi} = - a \Gamma \rho_{\varphi},
\end{equation}
where  $\Gamma$ is the decay width associated with an inflaton decay process.

In this work our interest in reheating lies in the contribution of the reheating phase to the overall redshift at which the energy density of the fluid $\rho_f$ becomes Planckian. This allows us to determine the range of physical scales at the initial time that corresponds to the range of CMB  observable  scales today. It turns out that we can write the ratio of the initial scale factor $a_i$ to today's scale factor  $a_0$ as
\begin{equation}\label{reratio1}
\frac{a_i}{a_0} = e^{-N_{\text{tot}}} \frac{a_{\text{end}}}{a_r}\frac{a_r}{a_0}.
\end{equation}
The above $N_{\text{tot}}$ is the total number of e-folds between $\rho_f = 1$ to the end of inflation, $a_{\text{end}}$ is the scale factor at the end of inflation, and $a_r$ is the scale factor at the beginning of the radiation dominated era. It is a straightforward computation to show that
\begin{equation}
\frac{a_{\text{end}}}{a_r}= \Big(\frac{\rho_r}{\rho_{\text{end}}}\Big)^{\frac{1}{3(1+\mathcal{W}_{re})}},
\end{equation}
where $\rho_r $ is the energy density of the radiation and $\rho_{\text{end}}$ is the total energy density at the end of inflation. Similarly, the ratio  $a_r/a_0$ is computed with the cosmological parameters provided by the flat $\Lambda$CDM model.

Here we shall restrict attention to the most trivial theory of reheating, i.e.  the instantaneous reheating, when constructing numerical examples in sections  \ref{numericscalar} and \ref{numerictensor}. For instantaneous reheating $a_{\text{end}}/a_r = 1$. However a  better motivated  model of reheating, such as the canonical reheating, has $a_{\text{end}}/a_r$ different than unity. In the canonical reheating model the inflaton decays to a single component non-relativistic matter. The duration of reheating in this case  is determined by the energy scale $\rho_{\text{end}}^{1/4}$ at the end of reheating. This energy scale is proportional to the so-called reheating temperature  $T_{\text{re}}$, which is the temperature  of the radiation fluid at the beginning of the radiation dominated era. When taking into account the constraints placed on inflationary models coming from reheating considerations, one needs to consider such effects for various $T_{\text{re}}$  [See \cite{kamion1,chiba} for a recent study on this subject]. Although as it turns out, a $T_{\text{re}}$ lower than the energy scale at the end of inflation (as in the canonical reheating case) only renders our predictions less interesting.

\section{Dynamics of the perturbations} \label{perturb}

\subsection{Fluid action}

In this section we derive an action principle for the coupled
fluid-inflaton-gravity perturbations.  We start by describing in more
detail our action principle for the fluid.

We describe the fluid flow by a set of comoving coordinates $x^{a} (\alpha^i,\lambda)$. Here $\alpha^i$ are a set of Lagrange coordinates chosen on an arbitrary spacelike hypersurface and $\lambda$ is an affine parameter, which together provide sufficient data to label a fluid flow line passing through a given point $x^{a}$ in spacetime. The flow of a perfect fluid is constrained by the conservation of the number density current along all flow lines,
\begin{eqnarray}\label{nd}
  \nabla_{a} J^{a} = 0.
\end{eqnarray}
The above $J^a$ is the number density current of the fluid defined as  $ n u^{a}$, with $u^{a}$ being the fluid four-velocity and $n$ the fluid number density. One then defines the fluid four-velocity by $dx^{a}/d \lambda \ (- dx^{a}/d \lambda \ dx_{a}/d \lambda)^{-1/2} $, thereby arriving at \cite{mukhanov}
\begin{equation} \label{j}
J^{a} =   \frac{\mathcal{F}(\alpha^i)}{ \sqrt{-g} \mathcal{J}} \frac{dx^{a}}{d\lambda},
\end{equation}
where $\mathcal{F}(\alpha^i)$ is some function of the Lagrange coordinates and $\mathcal{J} \equiv \mathcal{D}(x^{a})/\mathcal{D}(\alpha^i, \lambda)$ is the Jacobian associated with the coordinate transformation from the Lagrangian coordinates to the comoving coordinates. It can be easily seen that $\vec{J}$ defined in Eq. \eqref{j} solves the continuity equation \eqref{nd}.

With this description of the fluid flow, the general action for this model in the metric signature $(-+++)$ becomes
\begin{equation} \label{act00}
S=  \int  \sqrt{-g}\: d^4 x \: \Big[  \frac{R}{2 } +  \mathcal{L}_{m}\:\Big],
\end{equation}
where $g$ is the determinant of the spacetime metric, $R$ is the Ricci scalar, and $\mathcal{L}_m$ is the matter Lagrangian density. For convenience, we have set $\hbar = c = m_p = 1$, where $m_p$ is the reduced Planck mass defined as $(8 \pi G)^{-1/2}$. The matter Lagrangian density for this model is \footnote{One formulation of the Lagrangian density
of a perfect isentropic fluid is given by $$-\sqrt{-g} \rho_{f}\big (|\vec{J}| \big)  - \lambda \nabla_{a} J^{a},$$
where $\lambda$ is a Lagrange multiplier. If we use the Lagrangian fluid variables  explained on page 12,
then $\nabla_{a} J^{a} = 0$  by virtue of Eq. \eqref{nd}.
In this case, the fluid Lagrangian
density reduces to $- \sqrt{-g} \rho_{f}\big (|\vec{J}| \big) $. We refer the reader to Sec. 5 of \cite{brown} for
an extensive discussion of the Lagrangian fluid variables. The reader might
also find  various common formulations of the fluid action principle discussed on pages 31-36 of \cite{brown} insightful.}
\begin{equation}\label{lag0}
\mathcal{L}_m[g_{ab}, \varphi, x^{a}] = -\rho_{f}\big (|\vec{J}| \big) - \frac{1}{2} (\nabla \varphi)^2  - V(\varphi).
\end{equation}
The above $\rho_f$ is the energy density of the fluid and $\varphi$ is the inflaton field with a potential $V(\varphi)$.   Also $(\nabla \varphi)^2 $ is a shorthand for $ g_{ab}\nabla^{a}\varphi \nabla^{b} \varphi $ as usual. Note that the fluid  pressure $p_f$  can be defined in terms of the fluid  energy density $\rho_f$ via \cite{brown}
\be
p_f \equiv n \frac{\partial \rho_f}{\partial n} - \rho_f.
\ee
Alternatively, a fluid with an equation of state parameter $w$ has its energy density and pressure related by $p_f = w \rho_f$.

\subsection{Scalar perturbations}\label{perturbation}
In this section we will derive the linear equations of motion for the gauge invariant variables  associated with the scalar perturbations to the background fluid-inflaton model introduced in Sec. \ref{introduce}. It is convenient to derive these equations by expanding the action \eqref{act00} to second order in the scalar perturbations. Working with the action has privileges over the perturbed linearized Einstein equations. The clarity of the process of integrating out the non-dynamical components of the metric as well as the quantization of the perturbations are the main advantages of working with the action principle.

The scalar perturbations to the FRW metric in an arbitrary gauge can be written as \cite{baumann}
\begin{equation}\label{pertmetr}
ds^2 = - a(\eta)^2 \Big[ (1+ 2 \phi) d\eta^2- 2 B_{,i} dx^i d\eta - [(1-2 \psi) \delta_{ij} + 2 E_{,ij} ] dx^i dx^j \Big],
\end{equation}
for some scalar function $\phi$, $\psi$, $B$, and $     E$.  Similarly we write the perturbed scalar field as
\begin{equation} \label{pertinf}
\varphi(\eta,x^i) = \varphi(\eta) + \delta \varphi(\eta,x^i).
\end{equation}
We now insert the expansions \eqref{pertmetr} and \eqref{pertinf} of the metric and inflaton into the action \eqref{act00}, and expand to second order in the perturbations. The first order terms must vanish identically when the background variables are on-shell, so we drop all the first order terms. The second order gravitational and inflaton terms are
\begin{eqnarray} \label{111}
&&\delta_2 S_{\text{gravity}} \nonumber\\
&&=  \int d\eta \ \int \ d^3x \: \frac{a^2}{2} \Big\{ -6 \psi^{\prime 2} - 12 \mathcal{H}(\phi +\psi) \psi^{\prime} -9 \mathcal{H}^2  (\phi+\psi) ^2 -2 \nabla \psi . (2 \nabla\phi -\nabla \psi) \nonumber\\
&& -4 \mathcal{H} (\phi+\psi) \nabla^2 (B- E^{\prime})+ 4 \mathcal{H} \psi ^{\prime} \nabla^2 E- 4 \psi^{\prime} \nabla^2 (B-E^{\prime}) -4 \mathcal{H} \nabla\psi . \nabla B\nonumber\\
&& +6 \mathcal{H}^2 (\phi+\psi) \nabla^2 E - 4 \mathcal{H} \nabla^2 E \nabla^2 (B-E^{\prime}) + 4 \mathcal{H} \nabla^2 E \nabla^2 B +3 \mathcal{H}^2 (\nabla^2 E)^2 \nonumber\\
&&+ 3 \mathcal{H}^2 (\nabla B)^2 \Big \},
\end{eqnarray}
and
\begin{eqnarray} \label{112}
&& \delta_2 S_{\text{inflaton}} \nonumber\\
&&=  \int d \eta \ \int \ d^3 x \  \frac{a^2}{2} \Big\{ \big(a^2 V(\varphi)- \frac{\varphi^{\prime2}}{2}\big) \big ( \phi^2 - 3 \psi^2 - (\nabla B)^2 +(\nabla^2 E)^2 + 2 \psi \nabla^2 E +6 \phi \psi  \nonumber\\
&&- 2 \phi \nabla^2 E \big )-2 \big ( \phi - 3 \psi + \nabla^2 E \big) \big(a^2 V_{,\varphi}(\varphi) \: \delta\varphi + \varphi^{\prime} [\phi \varphi^{\prime} - \delta \varphi^{\prime}] \big) - \big[ (\nabla \delta \varphi)^2  \nonumber\\
&&+ 2 \varphi^{\prime} \nabla B . \nabla \delta \varphi+ a^2 V_{,\varphi \varphi}(\varphi) \delta \varphi ^2 - 4 \phi^2 \varphi^{\prime 2} + \varphi^{\prime 2} (\nabla B)^2 + 4 \phi \varphi^{\prime} \delta\varphi^{\prime} - (\delta \varphi^{\prime})^2 \big] \Big\},\nonumber\\
\end{eqnarray}
where $\nabla^2 \equiv \delta^{ij} \partial_i \partial_j$ and $\nabla A \ . \nabla B \equiv \delta^{ij} A_{,i} B_{,j}$.

Perturbing the fluid Lagrangian is a more delicate task. Note that the energy density of the fluid $\rho_f$ itself is not a dynamical variable. For a perfect isentropic fluid, the energy density is a function of the fluid number density $n$. One needs to perturb the fluid Lagrangian by perturbing $n$, and use \eqref{nd}  to relate the perturbations in $n$ to the metric perturbations and the perturbations in the fluid four-velocity, the latter being related to the perturbations of the comoving  frame of the fluid. Following \cite{mukhanov}, we denote the perturbations to the comoving frame of the fluid by a shift vector $\xi^i(\eta,x^j)$. Since we are interested in scalar perturbations here, we define $\xi_{i} \equiv \partial_i \xi$ for some function $\xi$. In short, we are writing the number density perturbations, $\delta n (\eta, x^i)$, in terms of $\xi (\eta, x^i)$ and other metric perturbation functions. The details of this calculation can be found in \cite{mukhanov}. The final result is
\begin{eqnarray} \label{113}
&& \delta_2 S_{\text{fluid}} \nonumber\\
&&= \int d \eta \ \int \ d^3 x \  a^4 \Bigg\{ \frac{1}{2} \rho^{0}_f \phi^2 +p^0_f \bigg( \frac{3}{2} \psi^2
-3 \phi \psi + \phi \nabla^2 E -\psi \nabla^2 E + \frac{1}{2} (\nabla^2 E)^2 - E_{,ij} E_{,ij} \nonumber\\
&& + \frac{1}{2} (\nabla B)^2 \bigg )  + (\rho^{0}_f +p^0_f) \bigg(\frac{1}{2} \xi^{\prime}_{,i} \xi_{,i} ^{\prime} + B_{,i} \xi_{,i} ^{\prime} + \phi \nabla^2 \xi  \bigg)  \nonumber\\
&&- \frac{1}{2} c_s ^2 (\rho^{0}_f+p^0_f) \big(3 \psi - \nabla^2 E - \nabla^2 \xi \big)^2 \Bigg\},
\end{eqnarray}
where $c_s$  is the sound speed  of the fluid defined by $c_s ^2 = p_f ^{\prime} / \rho_f ^{\prime}$.
Adding the contributions \eqref{111} to \eqref{113} and using the background equations of motion \eqref{fried1} and \eqref{fried2} together with integrations by parts in time and space, the total variation in the action simplifies to
\begin{eqnarray} \label{actionsimple}
&& \delta_2 S = \int d \eta \ \int \ \ d^3 x \ \frac{a^2}{2} \Big \{ -6 \big( \psi^{\prime 2} + 2 \mathcal{H} \phi \psi^{\prime} + \mathcal{H} ^2 \phi^2 \big ) - 4 \big( \psi^{\prime} + \mathcal{H} \phi \big) \nabla^2 (B- E^{\prime})  \nonumber\\
&&-4 \nabla \psi . \nabla \phi + 2 (\nabla \psi)^2 + a^2 (\rho^{0}_f +p^0_f) (\xi'_{,i} +B_{,i})^2 + a^2 (\rho^{0}_f +p^0_f)  (2 \phi \nabla^2 E - 6 \phi \psi +2 \phi \nabla^2 \xi ) \nonumber\\
&&- c_s ^2 a^2 (\rho^{0}_f +p^0_f) (3 \psi - \nabla ^2 E - \nabla^2\xi  )^2  + \varphi ^{\prime 2} \phi^2 +\frac{2}{a^2}(\phi - 3 \psi + \nabla^2 E) (\varphi^{\prime} \delta \varphi a^2 )^{\prime}-(\nabla \delta \varphi)^2  \nonumber\\
&& -2 \varphi^{\prime} \nabla B . \nabla \delta \varphi - a^2 V_{,\varphi \varphi}(\varphi) \delta \varphi^2 -4 \phi \varphi^{\prime} \delta \varphi^{\prime}+ (\delta \varphi^{\prime})^2 \Big \}.
\end{eqnarray}
In Eq. \eqref{actionsimple} six perturbation functions appear; four associated with the metric, one associated with the scalar field, and one associated with the fluid. It is convenient to replace the fluid perturbation function $\xi$ with a three-velocity potential defined as \cite{mukhanov}
\begin{equation}\label{deft}
\theta \equiv \frac{2 \sqrt{\beta} a}{c_s} (\xi' + B)
\end{equation}
which is related to the perturbation of the four-velocity  $\delta u ^{i}$ of the fluid by
\begin{equation}
\theta_{,i} =  -  \frac{2 \sqrt{\beta} a^2}{c_s} \delta u_{i}.
\end{equation}
In Eq. \eqref{deft} $\beta$ is defined as $ (a^2/2) (\rho^{0}_f + p^0_f) $. Using the definition \eqref{deft} and the constraint
\begin{equation}\label{vary}
\big(\frac{a c_s}{2} \sqrt{\beta} \theta \big) ^{\prime} = -a^2 c_s ^2 \beta (3 \psi - \nabla^2 E - \nabla^2 \xi + \frac{\phi}{c_s ^2} )
\end{equation}
obtained from varying the action \eqref{actionsimple} with respect to $\xi$, we eliminate all $\xi'$ and $\nabla^2 \xi$ appearing in Eq. \eqref{actionsimple} in favor of $\theta$  and the other metric perturbation functions. The final result is

\begin{eqnarray}\label{act2}
&& \delta_2 S = \int d \eta  \ \int \ d^3 x \   \frac{1}{2} \Bigg \{ -6 a^2 (\psi^{\prime} + \mathcal{H} \phi )^2 + 2 \frac{\beta a^2}{c_s ^2}\phi^2 -2 a^2 \nabla \psi . (2 \nabla \phi - \nabla \psi) \nonumber\\
&&+\frac{1}{2} \frac{(c_s a \sqrt{\beta} \theta)^{\prime 2}}{a^2 \beta c_s ^2} - \frac{1}{2}c_s^2 (\nabla \theta)^2 -6 c_s a \sqrt{\beta} \theta \psi^{\prime} - 2 a \sqrt{\beta} c_s \theta \bigg(\frac{\phi}{c_s ^2}\bigg)^{\prime} -4 a^2 \nabla^2 (B- E^{\prime})\nonumber\\
&&\bigg(\psi^{\prime} + \mathcal{H} \phi - \frac{1}{2} \varphi ^{\prime} \delta \varphi + \frac{c_s}{2 a} \sqrt{\beta} \theta \bigg)+ a^2 \varphi ^{ \prime 2} \phi^2 +2 ( \phi- 3 \psi) (\varphi ^{\prime} \delta \varphi a^2 )^{\prime} -a^2 (\nabla \delta \varphi)^2  \nonumber\\
&&- a^4 V_{,\varphi \varphi} \delta \varphi ^2 - 4 a^2 \phi \varphi^{\prime} \delta \varphi^{\prime}+a^2(\delta \varphi ^{\prime})^2  \Bigg \}.
\end{eqnarray}
Note that the action \eqref{act2} depends on the perturbation functions $B$ and $E$ only through the combination $B-E'$. Varying the action \eqref{act2} with respect to $B- E^{\prime}$ gives
\begin{equation}\label{vary2}
\psi^{\prime} + \mathcal{H} \phi  = \frac{1}{2} \varphi^{\prime} \delta \varphi  - \frac{c_s}{2 a} \sqrt{\beta} \theta_.
\end{equation}
As can  be seen from Eq. \eqref{act2}, $\phi$ is non-dynamical and  it can be integrated out using Eq. \eqref{vary2}, i.e. $\phi$ can be eliminated in favor of $\psi$, $\theta$, and $\delta \varphi$. Observe that by doing this we eliminate $E$, $B$, and $\phi$. This way we are left with three dynamical fields; $\psi$, $\theta$, and $\delta \phi$. The action \eqref{act2} after integrating out $B$, $E$, and $\phi$ becomes
\begin{eqnarray}\label{act3}
&& \delta_2 S = \int d \eta  \ \int \ d^3 x \   \frac{1}{2} \Bigg \{\theta^2 \Big[ -\frac{3}{2} c_s^2 \beta + \frac{\beta^2}{2 \mathcal{H}^2} - \frac{\beta' \mathcal{H}}{2 \mathcal{H}^2}+ \frac{\beta \mathcal{H}'}{2 \mathcal{H}^2}+ \frac{(c_s \sqrt{\beta} a)^{\prime 2}}{c_s^2 \beta a^2} - \frac{(c_s \sqrt{\beta} a)^{''}}{2 c_s \sqrt{\beta} a}  \nonumber\\
&&+a c_s \sqrt{\beta} \bigg(\frac{\sqrt{\beta}}{a \mathcal{H} c_s} \bigg)^{\prime}+ \frac{\varphi^{\prime 2}}{4 \mathcal{H}^2} c_s^2 \beta \Big]  + \theta \psi^{\prime} \Big( \frac{2 a \beta^{3/2}}{c_s \mathcal{H}^2}-6 c_s a \sqrt{\beta} - \frac{2(c_s \sqrt{\beta} a)'}{\mathcal{H} c_s ^2}+ \frac{a \varphi^{\prime2}}{\mathcal{H}^2} c_s \sqrt{\beta} \Big) \nonumber\\
&& + \theta \delta \varphi \Big( \frac{a^3}{\mathcal{H}} c_s \sqrt{\beta} V_{,\varphi} \Big) + \theta \delta \varphi' \Big( \frac{a}{\mathcal{H}} c_s \sqrt{\beta} \varphi' \Big) + \psi' \delta \varphi' \Big(\frac{2 a^2 \varphi'}{\mathcal{H}} \Big) + \psi' \delta \varphi \Big( \frac{2 a^4 V_{,\varphi}}{\mathcal{H}} + 6 a^2 \varphi' \Big) \nonumber\\
&& - \frac{2 \sqrt{\beta} a}{\mathcal{H} c_s} \theta' \psi' + \psi^{\prime 2} \Big( \frac{2 a^2 \beta}{c_s ^2 \mathcal{H}^2} + \frac{a^2 \varphi^{\prime 2}}{\mathcal{H}^2} \Big) + \frac{1}{2} \theta^{\prime2} + a^2 (\delta \varphi')^2 +  \frac{4 a^2}{\mathcal{H}} \nabla \psi . \nabla \psi' + 2 a^2 (\nabla \psi)^2 \nonumber\\
&&+ 2 \frac{a}{\mathcal{H}} c_s \sqrt{\beta} \nabla \psi . \nabla \theta - \frac{1}{2} c_s ^2 (\nabla \theta)^2  -a^2 (\nabla \delta \varphi)^2 + \delta \varphi ^2 \bigg[-a^4 V_{,\varphi \varphi} - \frac{a^4}{\mathcal{H}} \varphi' V_{,\varphi}\nonumber\\
&& + \bigg(\frac{a^2 \varphi^{\prime 2}}{2 \mathcal{H}} \bigg)' \bigg] \Bigg \}.
\end{eqnarray}

The functions $\delta \varphi$, $\psi$, and $\theta$  form the following two gauge-invariant variables
\begin{equation}\label{def1}
\nu = \frac{1}{\sqrt{2}} (\theta - 2 z \psi), \hspace{1 cm} \pi = a \delta \varphi + \zeta \psi,
\end{equation}
where $z= a \sqrt{\beta}/(\mathcal{H} c_s)$ and $ \zeta = a \varphi^{\prime}/\mathcal{H}$. The function $\nu$ is the gauge invariant fluid velocity potential, and the function $\pi$ is the gauge invariant inflaton perturbation.

So far in this computation we have not made any gauge specialization but kept the gauge arbitrary. One might imagine that this computation would be easier if one imposes some gauge conditions at the start, for example the Newtonian gauge conditions $B=E=0$. However, this gauge specialization is too restrictive and does not allow us to properly integrate out the non-dynamical variables. With that gauge choice, we would not have been able to derive the constraint equation \eqref{vary2} needed to integrate out $\phi$. Similar difficulties arise in the synchronous gauge where $\phi=B=0$ (but not in the inflaton comoving gauge where $\delta \varphi= E = 0$).

Rewriting the action \eqref{act3} in terms of $\nu$ and $\pi$ defined in Eq. \eqref{def1} gives \footnote{The action presented in this form is not applicable for the case of pressureless dust fluid. A redefinition of variables is necessary for this special case. } \footnote{For linearized scalar cosmological  perturbations, isentropic fluids admit a simple scalar field description. For the case of radiation fluid, one can replace the fluid Lagrangian density $-\rho_f$ with $  (\nabla \psi)^4 /4$. Cosmologies of scalar fields with non-canonical kinetic terms of this form had been previously studied (see e.g. \cite{kinflation}). }
\begin{equation}\label{act5}
 \delta_2 S =  \int d\eta \  \int \ d^3x  \ \Big \{ \mathcal{L}_{\text{fluid}}+\mathcal{L}_{\text{inflaton}}+ \mathcal{L}_{\text{int}} \Big\},
\end{equation}
where
\begin{equation}
\mathcal{L}_{\text{fluid}} = \frac{1}{2} \bigg( \nu^{\prime 2}- c_s ^2 (\nabla \nu)^2 + \Big[\frac{z^{\prime \prime}}{z}+4 A\Big] \nu ^2 \bigg)
\end{equation}
is the fluid Lagrangian density,
\begin{equation}
\mathcal{L}_{\text{inflaton}}=  \frac{1}{2} \bigg( \pi ^{\prime 2} - (\nabla \pi)^2 + \Big[\frac{\zeta ^{\prime \prime}}{\zeta}+ \frac{2}{\zeta^2}(4 z^2 A -(z B)^{\prime} ) \Big] \pi ^2 \bigg )
\end{equation}
is the inflaton Lagrangian density, and
\begin{equation}\label{interact}
\mathcal{L}_{\text{int}}=\Big (4 \sqrt{2} A \frac{z}{\zeta} - \sqrt{2} B \frac{\zeta^{\prime}}{\zeta^2}\Big ) \nu \pi + \sqrt{2} \frac{B}{\zeta} \nu \pi^{\prime}
\end{equation}
is the interaction Lagrangian density.
Here
\begin{eqnarray}
&&A\equiv \frac{\mathcal{H}^{\prime} }{4 \mathcal{H}^2} \varphi^{\prime 2}  - \frac{(\varphi^{\prime 2})^{\prime}}{8 \mathcal{H}} - \frac{\varphi^{\prime 2} \beta}{8 \mathcal{H}^2}(1-c_s ^2) + \frac{c_s ^{\prime}}{2 c_s} \frac{\varphi^{\prime 2}}{2 \mathcal{H}}
\end{eqnarray}
and
\begin{equation}
B\equiv \bigg(c_s - \frac{1}{c_s}\bigg) \frac{a \varphi^{\prime 2}  \sqrt{\beta}}{2 \mathcal{H}^2}
\end{equation}
are functions of the background variables. We remark here that the fluid influences the inflaton perturbations in three steps: first through the explicit interaction terms given in Eq. \eqref{interact}, second by modifying the expressions for the background functions $A$ and $B$ that involve the fluid variables, and third by modifying the background solutions which also influence the inflaton perturbations. Finally, observe that turning off the inflaton by setting $ \pi=0$ gives the standard gauge invariant Lagrangian density for the isentropic fluid \cite{mukhanov}, while turning off the fluid by setting $\nu=0$ gives the standard gauge invariant action for the inflaton perturbations \cite{sasaki1,sasaki2,mukhkuf}.

We can write the action \eqref{act5} more compactly as
\begin{equation}\label{actmat}
\delta_2 S = \int \ d\eta \ \int d^3x \ \frac{1}{2} \Big[\bs{\Pi}^{\prime \text{T}} \bs{\Pi}' - \nabla_i \bs{\Pi}^{\text{T}} \bs{c}^2 _{s} \nabla^{i} \bs{\Pi} + \bs{\Pi}^{\text{T}} \bs{\Gamma} \bs{\Pi}  + \bs{\Pi}^{\prime \text{T}} \bs{\lambda} \bs{\Pi} \Big],
\end{equation}
where  the matrices are defined as
\begin{eqnarray} \label{matrices1}
&& \bs{\Pi} \equiv \begin{pmatrix}
                                        \pi \\
                                        \nu
                                        \end{pmatrix},
\hspace{1 cm}   \bs{c}^2 _{s} \equiv \begin{pmatrix}
                         1 & 0 \\
                        0 & c_s ^2
                        \end{pmatrix},
\hspace{1cm}   \bs{\lambda} \equiv \begin{pmatrix}
                                                        0 & \sqrt{2} \frac{B}{\zeta} \\
                                                        - \sqrt{2} \frac{B}{\zeta}&0
                                                        \end{pmatrix},  \nonumber\\
&& \bs{\Gamma} \equiv \begin{pmatrix}
                                        \frac{\zeta''}{\zeta} + \frac{2}{\zeta} \big[ 4 z^2 A - (z B)' \big] & 4 \sqrt{2} A \frac{z}{\zeta}-\frac{\sqrt{2}}{2} B \frac{\zeta'}{\zeta^2}-\frac{\sqrt{2}}{2} \frac{B'}{\zeta} \\
                                        4 \sqrt{2} A \frac{z}{\zeta}-\frac{\sqrt{2}}{2} B \frac{\zeta'}{\zeta^2}-\frac{\sqrt{2}}{2} \frac{B'}{\zeta} & \frac{z''}{z}+4A
\end{pmatrix}.
\end{eqnarray}
We have suppressed the dependences on $\eta$ for all the background functions above. The Euler-Lagrange equation of motion for this action is
\be \label{mateqm}
\bs{\Pi}''- \big[ \bs{c}_s ^2 \nabla^2 + \bs{\Gamma}-\frac{1}{2}\bs{\lambda}' \big] \bs{\Pi}+ \bs{\lambda} \bs{\Pi}'=0.
\ee
Defining the Fourier modes of the perturbation functions using
\be
X(\bs{k},\eta) \equiv \int \ \frac{d^3x}{(2 \pi)^{3/2}} X(\bs{x}, \eta) e^{-i \bs{k}. \bs{x}},
\ee
 the corresponding equations of motion for each mode become
\begin{subequations}
\begin{eqnarray} \label{eqm}
&&\nu^{\prime \prime}  + \Big(c_s ^2 k^2- \frac{z^{\prime \prime}}{z}- 4 A \Big )\nu - \Big ( 4 \sqrt{2} A \frac{z}{\zeta} - \sqrt{2} B \frac{\zeta^{\prime}}{\zeta^2} \Big) \pi - \sqrt{2} \frac{B}{\zeta} \pi^{\prime}=0, \\
&&\pi^{\prime \prime}  + \Big[ k^2 -\frac{\zeta^{\prime \prime}}{\zeta} - \frac{2}{\zeta^2} (4 z^2 A - (z B)^{\prime}) \Big] \pi +\Big (\sqrt{2} \frac{B^{\prime}}{\zeta}- 4 \sqrt{2} A \frac{z}{\zeta}\Big ) \nu + \sqrt{2} \frac{B}{\zeta} \nu^{\prime}=0.
\end{eqnarray}
\end{subequations}
We have omitted $\nu$ and $\pi$ dependences on $\eta $ and $k$.
This is a system of coupled oscillators with time dependent coefficients.

\subsection{Tensor perturbations}\label{tperturbation}
To derive the linear equations of motion for the gravitational waves produced in this model, we expand the first term of the action \eqref{act00} to second order in the tensorial metric perturbations. Unlike  the scalar perturbations, the metric tensorial perturbations do not couple to the matter perturbations. The tensor perturbations to the metric are defined by \cite{weinberg}
\begin{equation} \label{metrict}
ds^2 = - a(\eta)^2 \big[d\eta^2 - (1+h_{ij}) dx^i dx^j \big],
\end{equation}
where $h_{ij}$ is a traceless and transverse matrix, i.e.  $\delta^{ij} h_{ij} = \partial _{i} h_{ij} = 0$.  It is not difficult to see that the tensor perturbations defined as such are gauge invariant \cite{weinberg}. We can express $h_{ij}$ as
\be \label{tensorpol}
h_{ij}(\bs{x},\eta) = \sum_{A = \{+,\times \}} \int \ \frac{d^3 k}{(2 \pi)^{3/2}}  \ \bigg[ e^{i \bs{k}. \bs{x}} \frac{2 \bar{h} (\bs{k},\eta)}{a(\eta)} e^{A} _{ij}(\hat{\bs{k}}) + \text{c.c.} \bigg],
\ee
where $A$ denotes the two independent polarization tensors $e^{A} _{ij}$ which are traceless and transverse and satisfy  $e^{A} _{ij} e^{B \ ij}= 2\delta^{AB}$.

The well known tensor perturbation action is given by \cite{baumann}
\begin{equation}\label{acttensor}
\delta_2 S_{\text{tensor}} =  \frac{1}{8}\int \ d \eta \ \int \ dx^3  a^2 \Big[ h_{ij} ^{\prime 2} - (\nabla h_{ij} )^2\Big].
\end{equation}
The equation of motion for the gravitational waves in the Fourier momentum space is given by
\begin{equation}\label{tensoreqm}
\bar{h}^{'' A} (k,\eta)+ \omega^2 (k,\eta) \bar{h} ^{A} (k,\eta)=0,
\end{equation}
with $\omega^2 (k) \equiv k^2 - a''/a$. This equation is shared in all models where the matter content is minimally coupled to gravity and the gravitational field action is the Einstein-Hilbert term. What renders  the evolution of the gravitational waves different in various models is the difference in  the background dynamics, i.e. the difference in $a''/a$ ratio. From the two Friedman equations \eqref{fried1} and \eqref{fried2} we have
\begin{equation}
\frac{a''}{a} = \frac{a^2}{6} \big( \rho- 3 p \big) = \frac{a^2}{6} \Big [ \big(\rho_{\varphi} - 3 p_{\varphi}) + \rho_f \big(1- 3 w \big) \Big].
\end{equation}
For the case of radiation ($w=1/3$) coupled to inflaton, the second term in the bracket vanishes and $a''/a$ is given by the same expression as in the case of the inflaton-only model. However, the background geometry and the dynamics of the inflaton field $\varphi$ is still influenced by the presence of the radiation fluid. We will explore these differences in Sec. \ref{numerictensor} where we numerically compute the spectrum of the primordial gravitational waves in the fluid-inflaton model.

\begin{figure}[h]
\caption{Below is the plot of $a''/a$ for the fluid-inflaton model (blue) and the inflaton-only model (dashed purple) with the $\varphi^2$ potential. The initial values in the inflaton-only model are chosen such that the background dynamics becomes identical to the fluid-inflaton model after the second transition. }
\centering
\includegraphics[width=0.8\textwidth]{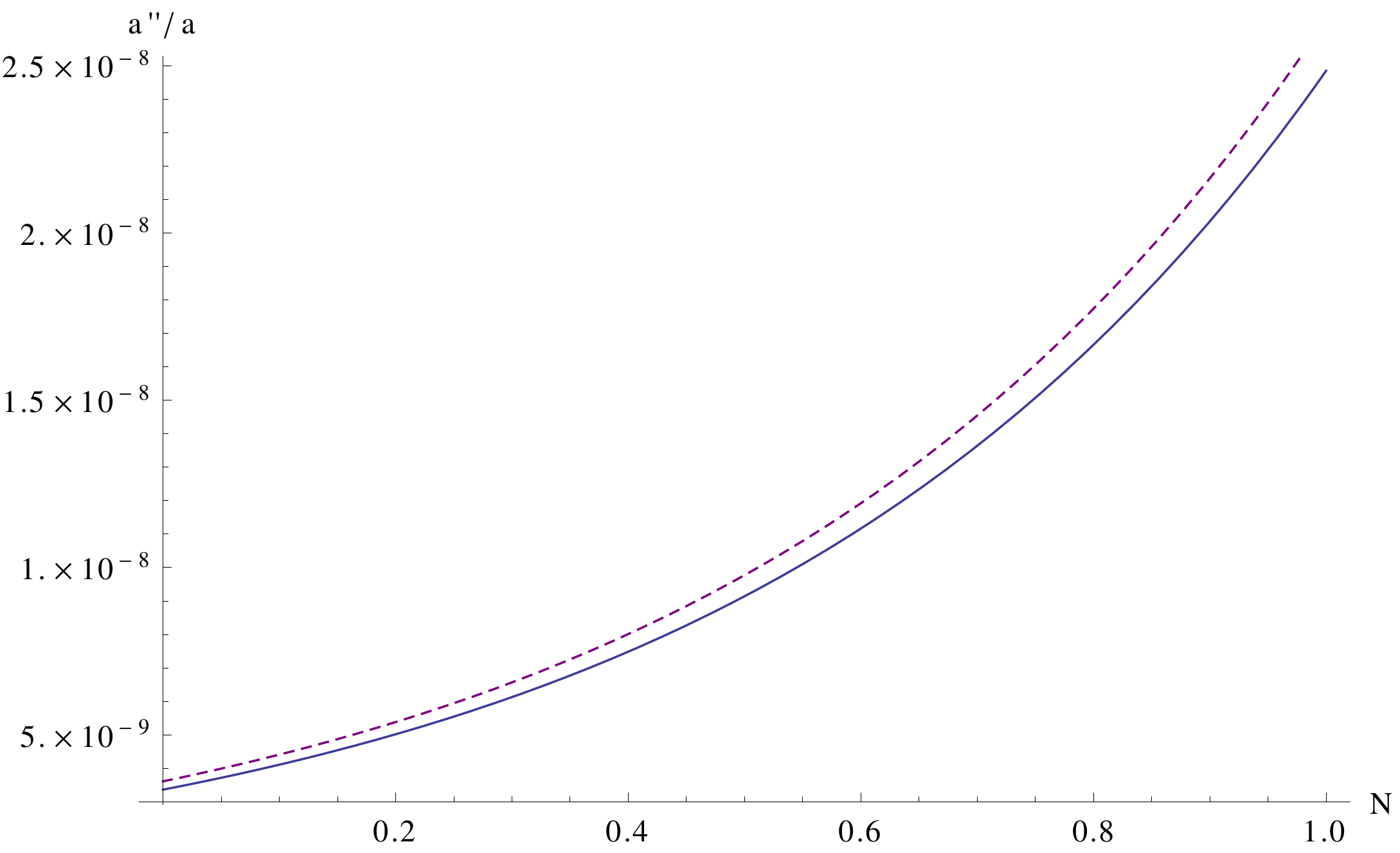}
\end{figure}

\section{Past and future decoupling of the scalar perturbations}\label{properties}
An interesting property of the fluid and inflaton scalar perturbations in this model is that they are only effectively coupled while the background energy densities are within a few orders of magnitude of one another. To establish this, we must examine the equations of motion \eqref{mateqm} both at early times and at later times when the background dynamics converges to that of single field inflation. The equations of motion \eqref{mateqm} are of the form (after Fourier transforming to the momentum space)
\begin{equation}\label{eqmmatrix}
\Pi_i ^{''} +\big[ \Omega_{ij} + \frac{1}{2} \lambda' _{ij} \big] \Pi_{j} + \lambda_{ij} \Pi'_{j}=0,
\end{equation}
where $\bs{\Pi}$ and $\bs{\lambda}$ are defined in Eq. \eqref{matrices1}, and
 \begin{eqnarray}\label{matrix}
&&
\mathbf{\Omega} \equiv - \bs{c}_s ^2 k^2 - \bs{\Gamma} =  \begin{pmatrix}
 k^2 - \frac{\zeta''}{\zeta} - \frac{2}{\zeta^2} \big[ 4 z^2 A - (z B)' \big] & -4 \sqrt{2} A \frac{z}{\zeta} + \frac{\sqrt{2}}{2} B \frac{\zeta'}{\zeta^2} +\frac{\sqrt{2}}{2} \frac{B'}{\zeta}\\
 -4 \sqrt{2} A \frac{z}{\zeta}  + \frac{\sqrt{2}}{2} B \frac{\zeta'}{\zeta^2}+\frac{\sqrt{2}}{2} \frac{B'}{\zeta} &  c_s ^2 k^2 -\frac{z''}{z} - 4 A
\end{pmatrix} .
\end{eqnarray}
As can be seen from Eqs. \eqref{eqmmatrix} and \eqref{matrix}, $\Omega_{11}$ and $\Omega_{22}$ can be interpreted as the effective masses of the inflaton and fluid in this model. Turning off the interactions, they become
\begin{equation}
\Omega_{11} \rightarrow k^2 - \zeta''/\zeta \hspace{1 cm}  \text{and}  \ \ \Omega_{22}\rightarrow c_s ^2 k^2 - z''/z
\end{equation}
as they would be in inflaton-only and fluid-only models. From Eq. \eqref{eqmmatrix}, one expects that the perturbations decouple  once
\begin{equation} \label{decoupling}
\lambda'_{12} , \ \Omega_{12}+\frac{1}{2} \lambda' _{12}  \ll  \ \Omega_{11}, \Omega_{22}.
\end{equation}
We can check this condition using the asymptotic background solutions obtained in Sec. \ref{backgroundanalytic}.  The result for  the distant past limit is given in Table. \ref{tab1}.

\begin{table} [h]
\begin{center}
    \begin{tabular}{| c | c | c |}
    \hline
     & $\frac{-1}{3}  < w < \frac{1}{3} $ & $w = \frac{1}{3}  $ \\ \hline
    $ \Omega_{11}, \Omega_{22} $& $ c_s ^2 k^2 + (\text{constant}) \  \eta^{-2}+ ... $ & $ c_s ^2  k^2 + (\text{constant}) \    \eta^{2} + ...           $             \\ \hline
   $ \Omega_{12}+ \frac{1}{2}\lambda' _{12} \ (i \neq j)$ & $\mathcal{O} \Big(\eta ^{\frac{8+6w}{1+3w}}  \Big)$ & $\mathcal{O} \Big(\eta ^{\frac{8+6w}{1+3w}}  \Big) $      \\ \hline
      $\lambda' _{12}  $ &   $\mathcal{O} \Big(\eta^{\frac{4}{1+3w}}  \Big)$ & $ \mathcal{O} \Big(\eta ^{\frac{4}{1+3w}}  \Big)$  \\ \hline
    \end{tabular}
\caption{Distant past forms of the background functions appearing in Eq. \eqref{eqmmatrix}.}
\label{tab1}
\end{center}
\end{table}
This calculation shows that the perturbation modes decouple in the limit $\eta \rightarrow 0$. Another question that could be asked is: does a mode of a given wavenumber $k$ decouple before the Planck scale is reached? To address this question, one could numerically compute the dimensionless ratio
\begin{equation}
\frac{\Omega_{12}+ \frac{1}{2} \lambda' _{12} }{\sqrt{\Omega_{11} \Omega_{22}}}
\end{equation}
just as $\rho_f$ becomes Planckian, and see whether it is small or large compared to unity.

Next we would like to show that this dynamical decoupling of the scalar perturbations also occur when the background dynamics begins to look like single field slow-roll inflation, i.e. when $\rho_f \lesssim \varepsilon_{\varphi} \eta_{\varphi} H^2$. This can be done using the late time solutions of the background equations \eqref{futureperturb}. The result is provided in Table. \ref{tab2}.
\begin{table}[h]
\begin{center}
    \begin{tabular}{| c | c |}
    \hline
      & $w = \frac{1}{3}$  \\ \hline
     $\Omega_{11} - k^2 $& $-\frac{2}{\Delta \eta^2}+\mathcal{O}(\bar{\varepsilon} _{\varphi}) $ \\ \hline
       $\Omega_{22} - c_s ^2 k^2 $ & $\big(\frac{2}{\Delta \eta^2} - 4 \Delta \eta + 2 \Delta \eta^4 \big) \bar{\varepsilon} _{\varphi} + \mathcal{O}(\bar{\varepsilon} ^2 _{\varphi} ) $\\ \hline
 $\Omega_{12} + \frac{1}{2} \lambda_{12} ^{\prime} $ & $\big( -\frac{1}{\sqrt{6}} + \frac{28}{\sqrt{6}} \Delta \eta ^3 \big) \bar{\varepsilon}^{3/2} _{\varphi} + \mathcal{O}(\bar{\varepsilon}_{\varphi} ^2) $ \\ \hline
$\lambda_{12} ^{\prime} $ & $\big( \frac{2}{\sqrt{6}} - \frac{8}{\sqrt{6}} \Delta \eta ^3 \big) \bar{\varepsilon}^{3/2} _{\varphi} + \mathcal{O}(\bar{\varepsilon}_{\varphi} ^2)  $   \\ \hline
    \end{tabular}
\caption{Future forms of the background functions appearing in Eq. \eqref{eqmmatrix}.}
\label{tab2}
\end{center}
\end{table}
From this we  see that early on during the second transition we have
\begin{eqnarray}\label{futuredecoupling}
&& \Omega_{11}-k^2 \sim \Delta \eta ^{-2} \sim 1 \gg \Omega_{12}+ \frac{1}{2} \lambda^{\prime} _{12} \sim \lambda^{\prime} _{12} \sim \bar{\varepsilon} ^{3/2} _{\varphi} ,\nonumber\\
&& \Omega_{22}- c_s ^2 k^2 \sim\mathcal{O}( \bar{\varepsilon} ^2 _{\varphi}) \lesssim \Omega_{12}+ \frac{1}{2} \lambda^{\prime} _{12} \sim \lambda^{\prime} _{12} \sim  \bar{\varepsilon} ^{3/2} _{\varphi},
\end{eqnarray}
where we have used $\Delta \eta =1 $ at the second transition consistent with our choice of scale factor normalization  in Sec. \ref{backgroundanalytic}. From Eq. \eqref{futuredecoupling} we see that the inflaton mode has effectively decoupled from the fluid mode by the second transition. However, this does not appear to be the case for the fluid mode. It is in fact simple to show analytically that
\be
 \Omega_{22}- c_s ^2 k^2=\frac{z''}{z}+4 A \rightarrow 0 , \hspace{1cm} \text{in the limit of } a \rightarrow \sqrt{\frac{3}{\Lambda_1}} \frac{1}{\Delta \eta},
\ee
i.e., the effective mass of the fluid becomes equal to its wavenumber squared in the limit of the background geometry becoming de Sitter. Thus, when considering
the issue of the decoupling of  the fluid modes at the second transition, one must consider the full fluid effective mass $\Omega_{22}$ instead. As we will see
in the next subsection, we shall be primarily interested in the scalar modes with wavenumbers $k \gtrsim k_{\text{min}}$. Using the value of $k_{\text{min}}$ obtained in Sec. \ref{timescale}, we have
\be
\Omega_{22} \gtrsim k_{\text{min}} ^2 \approx (\mathcal{E}_0 \Lambda_0)^{1/2} \sim \bar{\varepsilon}_{\varphi} \gg  \Omega_{12}+ \frac{1}{2} \lambda^{\prime} _{12} \sim \lambda^{\prime} _{12} \sim  \bar{\varepsilon} ^{3/2} _{\varphi},
\ee
where we used $k_{\text{min}} \sim (\mathcal{E}_0 \Lambda_0)^{1/4}$ and $\mathcal{E}_0 \sim \bar{\varepsilon} ^{2} _{\varphi}/\Lambda_0$ consistent with  our choice of $\Delta \eta=1$ at the second transition (also note that we are assuming $\Lambda_0 \approx \Lambda_1$ for simplicity). Therefore, it appears that the  fluid and inflaton perturbation modes of primary interest will eventually decouple again early on during the second transition.

\subsection{Adiabaticity of the scalar perturbations ; $w = \frac{1}{3}$}\label{vacuum}

 The analysis done in Sec. \ref{properties} indicates that the fluid and inflaton modes dynamically decouple at early times, reducing  to two uncoupled harmonic oscillators with time dependent frequencies.  We now show that if the fluid is radiation ($w=1/3$), both the fluid and inflaton modes are in an adiabatic regime at early times, that is the effective frequencies $\sqrt{\Omega_{ii}}$ of those modes evolve in timescales much longer than the frequencies themselves. This concept of adiabaticity replaces and generalizes the usual concept of modes being "inside the horizon".   Indeed for both the fluid and inflaton modes, the effect of the background geometry on subhorizon perturbations diminishes as $\eta^2$ with $\eta \rightarrow 0$. This can be seen in Table. \ref{tab1}, where we showed
\begin{equation}\label{freq}
\Omega _{22} \equiv c_s ^2 k^2-\frac{z''}{z} -4A = c_s ^2 k^2 + \mathcal{O}(\eta^2), \hspace{1 cm} \Omega_{11} \equiv k^2 - \frac{\zeta''}{\zeta} - \frac{2}{\zeta^2} \big[4 z^2 A - (z B)' \big] =  k^2 + \mathcal{O}(\eta^{2}).
\end{equation}
This indicates
\begin{equation}\label{adab}
\bigg|\frac{\Omega'_{22}}{\Omega^{3/2}_{22}} \bigg|,  \bigg|\frac{\Omega'_{11}}{\Omega^{3/2}_{11}}\bigg|\ll 1
\end{equation}
as $\eta \rightarrow 0$. Therefore in this limit the scalar modes behave like simple, time-independent harmonic oscillator for which the general state furnishes a natural choice of initial state, a generalization of the usual Bunch-Davies vacuum.

Although all modes become adiabatic sufficiently early in time, not all modes will have become adiabatic by the time the Planck scale is reached, which is the earliest time our model can be trusted.  We look into this matter by computing the adiabaticity ratio \eqref{adab} evaluated for $k_{\text{min}}$ when the fluid becomes Planckian.  Using the background perturbative solutions found in Sec. \ref{backgroundanalytic}, we find (for $w=1/3$)
\begin{equation}
\bigg|\frac{\Omega'_{11}}{\Omega^{3/2} _{11}} (k_{\text{min}}) \bigg| \approx  \Lambda_0 ^{3/4}, \hspace{1 cm} \bigg|\frac{\Omega' _{22}}{\Omega^{3/2} _{22}} (k_{\text{min}}) \bigg| \approx 10 \ \Lambda_0 ^{1/4},
\end{equation}
where $\Lambda_0  = V(\varphi_0)$. Note that here we have specialized to the class of solutions for the inflaton field for which the field freezes at early times. Since the inflaton field is expected to be at the GUT scale during inflation, we find
\be
\bigg |\frac{\Omega'_{11}}{\Omega^{3/2} _{11}} (k) \bigg | \gtrsim 10^{-6} , \hspace{1cm} \bigg|\frac{\Omega'_{22}}{\Omega^{3/2} _{22}} (k) \bigg |  \gtrsim 0.1, \hspace{0.5 cm} \text{for all } \ k \lesssim k_{\text{min}}.
\ee
Evidently, the fluid modes with wavenumbers smaller than $k_{\text{min}}$ will not enjoy a strong adiabaticity at the Planck scale. We will therefore be interested in having the observable scales correspond to wavenumbers larger than $k_{\text{min}}$.

\subsection{Adiabaticity of the tensor perturbations ; $w = \frac{1}{3}$} \label{tvacuum}

Just as in the scalar perturbation case, the tensor modes are in an adiabatic regime at early times. For tensor perturbations to be in an adiabatic regime at early times, the frequency of the gravitational waves $\omega$ defined in Eq. \eqref{tensoreqm} must satisfy
\begin{equation}\label{tadab}
\bigg|\frac{\omega ' }{\omega ^2}\bigg| \ll 1.
\end{equation}
Using the background solution for the scale factor found in Sec. \ref{backgroundanalytic} one finds
\begin{equation}
\omega^2 (k) = k^2 - \frac{a''}{a} = k^2 + \mathcal{O} (\eta^2)
\end{equation}
indicating that $\omega^{2} (k) \rightarrow k^2$ for all tensor modes as $\eta \rightarrow 0$. However, we must see for what modes the conditions given in Eq. \eqref{tadab} are met by the time the energy density of the fluid becomes Planckian.  A short calculation similar to the one done for the scalar modes  shows
\begin{equation}
\frac{\omega' (k_{\text{min}})}{\omega^2 (k_{\text{min}})} \approx \Lambda_0 ^{1/4},
\end{equation}
where the expression is calculated when the fluid is at the Planck scale. Therefore for all gravitational waves with wavenumbers $k \gtrsim k_{\text{min}} $, there exists a natural choice of ground state at the initial time when the fluid energy density is Planckian.

\subsection{Non-adiabaticity of perturbations ; $w \neq \frac{1}{3}$} \label{vacuum1}
In the case of a fluid with an equation of state parameter $w \neq 1/3$,  a short calculation using the results of Sec. \ref{backgroundanalytic} gives
\begin{eqnarray}
&& \Omega_{22} = c_s ^2 k^2- \frac{2-6w}{(1+3w)^2} \eta^{-2} + \mathcal{O} (\Lambda_0 \eta^2), \hspace{1cm} \Omega_{11} = k^2 - \frac{2-6w}{(1+3w)^2} \eta^{-2}+ \mathcal{O} (\Lambda_0 \eta^2), \nonumber\\
&& \hspace{3cm} \omega^2 = k^2- \frac{2-6w}{(1+3w)^2} \eta^{-2} + \mathcal{O} (\Lambda_0 \eta^2).
\end{eqnarray}
Here one cannot achieve initial adiabaticity in the sense of \eqref{adab} for all modes of interest, particularly for modes near $k_{\text{min}}$ which are of primary interest in this paper.

The non-adiabaticity for a mode is an indication of its interaction with the background geometry. In such cases one no longer has a natural choice of ground state. One could take different approaches to parameterize such initial states as excited states, the core idea being to constrain such states using backreaction considerations of the energy density stored in these initial quanta. Computations using excited initial states are more speculative due to our ignorance of the correct initial occupation numbers for perturbation quanta. Therefore we refrain from constructing numerical examples for such scenarios in sections \ref{numericscalar} and \ref{numerictensor}.

\section{Curvature and entropy perturbations in this model}\label{iso}

In the cosmological theory of small fluctuations, conservation laws can exist for certain gauge invariant variables in some circumstances. The existence of conserved quantities is the primary reason why predictions of some inflationary models can be tested despite our ignorance of what went on immediately after inflation. Of primary interest in this regard is a gauge invariant function known as the comoving curvature perturbation $\mathcal{R}$  defined by
\be \label{R}
\mathcal{R}\equiv \psi - \mathcal{H} \frac{\delta q}{a(\rho^{0} + p^{0})} = \psi - \mathcal{H} \frac{\delta u}{a},
\ee
where $\delta q $ and $\delta u$ are the three-momentum and the three-velocity potentials associated with a fluid [see appendix \ref{ptensor}]. The comoving curvature perturbation gives the initial conditions for the temperature and polarization fluctuations that we observe today in the CMB \cite{weinberg}. Taking $\mathcal{R}$ to the Fourier space using $\mathcal{R}(\bs{x},\eta) = \int \ d^3k /(2 \pi)^{3/2} \  \mathcal{R}(\bs{k},\eta) e^{i \bs{k} .\bs{x}} $, one can show  \cite{baumann}
\be
\mathcal{R}^{\prime}(k,\eta) = - 3 \frac{\mathcal{H} p^{0 \prime}}{ \rho^{0 \prime}} \mathcal{S}(k,\eta)+ \mathcal{O}\bigg(\frac{k^2}{\mathcal{H}^2}\bigg)
\ee
where $\mathcal{S}(k,\eta)$ is the Fourier transform of a gauge invariant function called the entropy perturbation,
\begin{equation} \label{entropy}
\mathcal{S}\equiv \mathcal{H} \Big (\frac{\delta \rho}{\rho^{0 \prime}}- \frac{\delta p}{p^{0 \prime}} \Big ),
\end{equation}
where $\delta p $ and $\delta \rho$ are the pressure and energy density perturbations. This function is the gauge invariant measure of  the non-adiabaticity of scalar perturbations. What we are primarily interested in is whether the comoving curvature perturbation for the fluid-inflaton model is conserved on super horizon scales.

It is a remarkable fact that for any background cosmology, one can find solutions of the perturbation equations \eqref{mateqm} for which $\mathcal{R}(k)$ is conserved as long as it remains in the superhorizon regime,  i.e.  as long as $k \ll \mathcal{H}$ \cite{star1,star4,weinbergiso1,weinbergiso2}. This fact remains valid regardless of the specifics of all subsequent cosmological eras. These perturbations, known as the adiabatic perturbations, can be intuitively realized as local spatial curvature fluctuations inducing a time-delay effect in the background geometry \cite{hudodel}. All linear fluctuations in single component cosmological models are  known to be of this kind. On the other hand, for more complicated cosmological models, such as multifield inflationary models,  non-adiabatic solutions can exist. For these solutions, $\mathcal{S}(k)$ will generally not vanish in the superhorizon regime. Non-vanishing of $\mathcal{S}(k)$ causes   the comoving curvature perturbation $\mathcal{R}(k)$ to change even in the superhorizon regime. Without a quantity that remains conserved in the superhorizon regime, predictions of an inflationary model are only testable if one  chooses a specific theory of reheating.

Luckily, the entropy perturbations are unimportant in our model. We establish this fact by  showing that the entropy perturbation function $\mathcal{S}$ reduces to its single field counterpart after the second transition. Using the definition \eqref{entropy} we have
\be
\mathcal{S} =  \mathcal{H} \Big [\frac{\delta \rho_f + \delta \rho_{\varphi}}{\rho^{0 \prime}}- \frac{\delta p_f + \delta p_{\varphi}}{p^{0 \prime}} \Big ],
\ee
where $\delta \rho_{f (\varphi)}$ and $\delta p_{f (\varphi)}$ are the energy density and pressure perturbations of the fluid (inflaton), while $\rho^0$ and $p^0$ are the overall background energy density and pressure. A short computation using the Friedmann equations \eqref{fried1} and \eqref{fried2} gives
\be
\rho ' = -\frac{6 \mathcal{H}^3}{a^2} \varepsilon, \hspace{2cm} p' = \frac{2 \mathcal{H}^3}{a^2} \varepsilon (3-2 \eta),
\ee
where $\varepsilon \equiv - \dot{H}/H^2$ and $\eta \equiv - \ddot{H}/{(2 \dot{H} H)}$ are the two main slow-roll functions parameterizing the degree to which the background de Sitter symmetry is broken. It is also convenient to define
\bea \label{newslowroll}
&&\eta^0 _{\varphi} \equiv \frac{-1}{2 H^3} \frac{\ddot{H}_{\varphi}}{\varepsilon_{\varphi}}, \hspace{1cm} \eta^0 _{f} \equiv  \frac{-1}{2 H^3} \frac{\ddot{H}_{f}}{\varepsilon_{\varphi}} \nonumber\\
&& \varepsilon_{\varphi} \equiv - \frac{\dot{H}_{\varphi}}{H^2}, \hspace{1cm} \varepsilon_{f} \equiv - \frac{\dot{H}_{f}}{H^2},
\eea
where the subscript $\varphi (f)$ on derivatives of the Hubble parameter denotes the contribution due to the inflaton (fluid).

Next we examine the fluid and inflaton perturbation functions. The inflaton  perturbation functions are \cite{baumann}
\be \label{infpertx}
\delta \rho_{\varphi} = \delta X  + V_{,\varphi} \delta \varphi, \hspace{1cm} \delta p _{\varphi} = \delta X - V_{,\varphi} \delta \varphi,
\ee
where we defined $\delta X \equiv \varphi' \delta \varphi'/ a^2 - \varphi^{\prime 2} \phi  / a^2 $.
The fluid perturbation functions can be derived from the linearized Einstein equations \cite{mukhanov}
\be
\delta \rho_{f} = \frac{2 \beta}{a^2} [ 3 \psi - \nabla^2 E - \nabla^2 \xi], \hspace{1cm} \delta p_f = c_s ^2 \delta \rho_f,
\ee
which using the constraint equation \eqref{vary} reduces to
\be \label{fpertx}
\delta \rho_f = - \frac{2}{a^4 c_s ^2} \bigg[ \big(\frac{a c_s}{2} \sqrt{\beta} \theta \big)' + a^2 \beta \phi \bigg].
\ee
Using Eqs. \eqref{newslowroll} to \eqref{fpertx}, we write $\mathcal{S}$ as a power series in the fluid slow-roll parameters $\eta^0 _f $ and $\varepsilon_{f}$,
\be
 \mathcal{S} = \mathcal{S}^{(0)} + \mathcal{S}^{(1)} + ...
\ee
with
\be
\mathcal{S}^{(0)} = \frac{1}{\mathcal{H}}\frac{2 (\eta^0 _{\varphi}-3)}{3(3-2 \eta^0 _{\varphi})} \mathcal{R}' _{\varphi}
\ee
and
\bea \label{s1}
&& \mathcal{S}^{(1)} = \frac{1}{\mathcal{H}}\Bigg[ \frac{2 (\eta^0 _{\varphi} -3)}{3 (3-2 \eta^0 _{\varphi})\varepsilon_{\varphi}} \varepsilon_f - \frac{2 \eta^0 _{\varphi}}{(3-2 \eta^0 _{\varphi})^2} \bigg( \frac{\eta^0 _{f}}{\eta^0 _{\varphi}} - \frac{\varepsilon_{f}}{\varepsilon_{\varphi}} \bigg) \Bigg] \mathcal{R}'_{\varphi} - \frac{1}{\mathcal{H}} \Bigg[\frac{\varepsilon_f}{3 \varepsilon_{\varphi} c_s ^2} \bigg(1+ \frac{3 c_s ^2}{3-2 \eta^0 _{\varphi}} \bigg) \Bigg] \mathcal{R}'_f \nonumber\\
&&+ (\mathcal{R}_{\varphi} - \mathcal{R}_f) \Bigg[ \bigg( \frac{1}{3 c_s ^2}-\frac{2}{3} \bigg ) \varepsilon_f - \frac{2 \eta^0 _{\varphi}}{3-2 \eta^0 _{\varphi}} \bigg( \frac{\eta^0 _{f}}{\eta^0 _{\varphi}} - \frac{\varepsilon_{f}}{\varepsilon_{\varphi}} \bigg)\Bigg].
\eea
Here we have used the definitions
\be
\mathcal{R}_{\varphi} \equiv \frac{\pi}{\zeta} \hspace{1cm} \text{and} \hspace{1cm} \mathcal{R}_f \equiv- \frac{\nu}{\sqrt{2} z}
\ee
which are the comoving curvature perturbations in absence of the other species.
Judging from the slow-roll parameters appearing in Eq. \eqref{s1}, one can see that after the second transition when $\eta^0 _{f} \ll \eta^0 _{\varphi}$ and $\varepsilon_f \ll \varepsilon_{\varphi}$, $\mathcal{S}$ converges to $\mathcal{S}^{(0)}$, which is the entropy perturbation function of a single inflaton field. This suggests that as the fluid energy density becomes subdominant to that of inflaton, the entropy between the two species diminishes and the overall entropy perturbation function reduces to its value in the absence of fluid. This indicates that our model does not produce non-negligible entropy perturbations, just as in the case of single field inflation.

\section{Power spectrum of the scalar perturbations} \label{power}

The power spectrum of the comoving curvature perturbation is the primary tool by which predictions of inflationary models are experimentally tested. Using the definition \eqref{R} we can write
 \begin{equation}
 \mathcal{R} \equiv \psi - a \mathcal{H} \frac{\delta q_f + \delta q_{\varphi}}{\varphi^{ \prime 2} + 2 \beta}= \psi - \frac{\mathcal{H}}{\varphi^{\prime 2}+2 \beta}  \bigg[ - \varphi^{\prime} \delta \varphi + c_s \frac{\sqrt{\beta}}{a} \theta \bigg ]  = \frac{\varphi^{\prime 2} \frac{\pi}{\zeta}}{\varphi^{\prime 2}+2 \beta} -\frac{2 \beta \frac{\nu}{\sqrt{2}z}}{\varphi^{\prime 2} + 2 \beta},
 \end{equation}
where the sum of momentum perturbations $\delta q _f + \delta q _{\varphi}$ was obtained from the right hand side of Eq. \eqref{vary2}, being $-T^{0}_{i} /2$. Also the gauge invariant variables $\nu$ and $\pi$ were introduced in Eq. \eqref{def1}. Note that $\pi/\zeta$ and $\nu/\sqrt{2}z$ would be the comoving curvature perturbations in inflaton or fluid-only models.

Since we are interested in understanding how primordial quantum fluctuations seed the formation of CMB anisotropies, we need to compute the quantum correlation functions of the comoving curvature perturbation $\mathcal{R}$. The presence of derivative interactions between $\nu$ and $\pi$ in action \eqref{act5} renders the quantization procedure non-trivial. The details of this procedure is provided in appendix \ref{app2}.  It turns out that one can write the quantized modes of the comoving curvature perturbation $\mathcal{R}$ as
\begin{equation}\label{qR}
\hat{\mathcal{R}}_{\mathbf{k}} (\eta ) = \mathbf{W}^{T} (\eta) \hat{\bs{\Pi}}_{\mathbf{k}} (\eta).
\end{equation}
Here
\begin{equation}
\mathbf{W} \equiv \begin{pmatrix}
                                \frac{\varphi^{\prime 2}}{\varphi^{\prime 2}+ 2 \beta} \zeta^{-1} \\
                                \frac{-2 \beta }{\varphi^{\prime 2}+ 2 \beta}  (\sqrt{2} z)^{-1}
                                \end{pmatrix},
\end{equation}
and
the quantized modes for the generalized solutions $\hat{\bs{\Pi}}_{\mathbf{k}} (\eta)$ are defined as
\begin{equation}\label{qpi}
\hat{\bs{\Pi}}_{\mathbf{k}} (\eta) \equiv \mathbf{\Pi}_{A} (k,\eta) \hat{a}_{A, \mathbf{k}} + \mathbf{\Pi}^{*} _{A} (k, \eta) \hat{a}^{\dagger} _{A, -\mathbf{k}},
\end{equation}
where similar to the notation used in Sec. \ref{properties}
\begin{equation}
\mathbf{\Pi}_{A} \equiv \begin{pmatrix}
                                   \pi_{A} \\
                              \nu_{A}
                                \end{pmatrix}
\end{equation}
with the set $\{ \mathbf{\Pi}_1,  \mathbf{\Pi}_2, \mathbf{\Pi}^{*} _1, \mathbf{\Pi}^{*} _2 \}$ of four linearly independent solutions spanning the four dimensional space of solutions for the coupled system of differential equations  \eqref{mateqm}.  Also, the annihilation operators $\hat{a}_{A \mathbf{k}}$ eliminate the quantum mechanical ground states $|\Omega \rangle _{A}$ for the inflaton and the fluid perturbations. The existence of natural ground states is guaranteed by  virtue of the adiabaticity of the scalar modes at early times discussed in Sec. \ref{properties} for the case of inflaton coupled to radiation, which is our primary interest.

Given the fact that there is a natural choice of ground state for the inflaton and fluid perturbations, we compute the power spectrum of the comoving curvature perturbation in the initial state
\begin{equation}
|\mathcal{I} \rangle \equiv |\Omega_1 \rangle \otimes |\Omega_2 \rangle,
\end{equation}
where $|\Omega_{1 (2)} \rangle$ is the initial ground state for inflaton (fluid) perturbations.  The power spectrum of the comoving curvature perturbation then becomes
\begin{equation}
\langle \mathcal{I}| \hat{\mathcal{R}}^{\dagger} _{\mathbf{k}} (\eta_{\text{out}}) \hat{\mathcal{R}}_{\mathbf{k'}} (\eta_{\text{out}}) | \mathcal{I} \rangle \equiv   \delta^{3} (\mathbf{k}+\mathbf{k'}) \mathcal{P}_{\mathcal{R}} (k)
\end{equation}
where $\eta_{\text{out}}$ is conventionally taken to be the conformal time for which $k= \mathcal{H}$, and
\begin{eqnarray} \label{sspectrum}
&& \mathcal{P}_{\mathcal{R}}(k) =  \sum_{A} \Big( \mathbf{W}^{T}   \mathbf{\Pi}_{A}(k) \Big)\Big( \mathbf{W}^{T}   \mathbf{\Pi}^{*} _{A}(k) \Big)\nonumber\\
&& = \sum_{A} \Bigg[ \bigg (\frac{\varphi ^{\prime 2}}{\varphi ^{\prime 2}+ 2 \beta }\bigg)^2 \frac{|\pi_{A} (k)|^2 }{\zeta^2} + \bigg (\frac{2 \beta}{\varphi ^{\prime 2}+ 2 \beta }\bigg)^2 \frac{|\nu_{A} (k)|^2 }{2 z^2} - 2 \  \text{Re} \bigg \{ \frac{2 \beta \varphi^{\prime 2}}{(\varphi ^{\prime 2}+ 2 \beta)^2}  \frac{\pi_{A} (k) \nu^{*} _{A} (k)}{\sqrt{2} z \zeta} \bigg \} \Bigg], \nonumber\\
&&
\end{eqnarray}
where all the background and perturbation functions are evaluated at $\eta_{\text{out}}$. Note that the scalar power spectrum receives most of its contribution from the first term in the square bracket, since $\sqrt{\beta}/\varphi' \rightarrow 0 $ with the fluid energy density becoming subdominant to that of inflaton.

It is often useful to work with a dimensionless power spectrum $\Delta^2_{\mathcal{R}}$ defined as
\be \label{dlsspectrum}
\Delta^2_{\mathcal{R}} \equiv \frac{k^3}{2 \pi^2} \mathcal{P}_{\mathcal{R}}.
\ee
Since the standard picture of inflation predicts a nearly scale invariant power spectrum,  $\Delta^2 _{\mathcal{R}}$ is expressed using a simple ansatz
\be
\Delta^2 _{\mathcal{R}} = A_s \Big( \frac{k}{k_{*}} \Big) ^{n_s -1},
\ee
where
\be
A_s = \frac{1}{8 \pi^2} \frac{H^2}{\varepsilon}
\ee
 is the value of the power spectrum evaluated at a pivot scale $k_{*}$ and
\be
n_s -1 = 2 \eta -4 \varepsilon + \text{terms higher order in slow-roll},
\ee
 is a measure of the scale dependence of the scalar power spectrum \cite{baumann}.

\subsection{Case study} \label{numericscalar}

We now numerically evaluate the power spectrum \eqref{dlsspectrum} for
the case of radiation coupled to an inflaton with a $\varphi^2$
potential. As discussed previously, specializing to radiation has the
advantage of bringing about the adiabaticity of perturbation modes at
initial times.

As already discussed in Sec. \ref{introduce}, we shall fix the initial energy scale of the radiation at the Planck scale and specialize to the class of  background cosmology for which $\varphi$ approaches a constant value in the $\eta \rightarrow 0$ limit. In this case, the remaining free parameters are the initial value of the inflaton field and the coupling parameter of the inflaton potential $m^2 /2$. A super-Planckian field variation of order $\Delta \varphi \sim 15$ during inflation together with the measured values of   $\Delta^2 _{\mathcal{R}}$ and its scale dependence parameter $n_s$ require $ 10^{-6} \lesssim m \lesssim 10^{-5}$ \cite{baumann}. Knowing the range of these two parameters, we look for values that result in a suitable fit to  the reconstructed values of the amplitude $A_s$ and the scale dependence parameter $n_s$ for the $\varphi^2$ potential model at the pivot scale $ k_{*} = 0.05 \ \text{Mpc}^{-1} $, as reported by the Planck collaboration \cite{planck1}.

\begin{figure}[h]
\centering
\includegraphics[width=0.9 \textwidth]{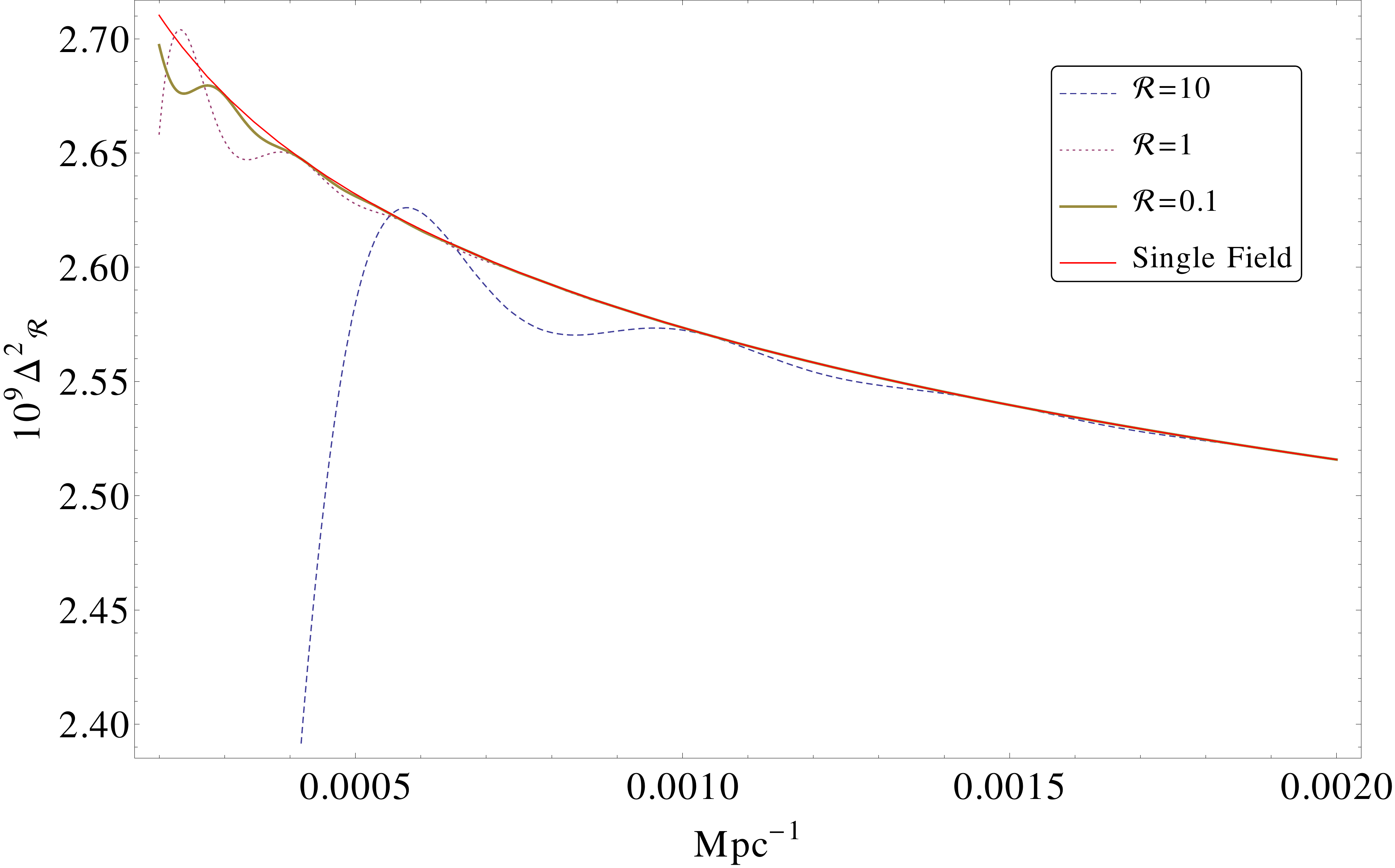}
\caption{The power spectrum $\Delta^2 _{\mathcal{R}}$ versus $k$ for radiation coupled
  to an inflaton with the potential $V(\varphi) = m^2
  \varphi^2/2$ for $2 \lesssim l \lesssim 28$ assuming instantaneous
  reheating.  The parameter $\mathfrak{R}$ is the ratio of the fluid
  energy density to the inflaton energy density when the $l=2$ mode
  left the horizon.}
\label{sspectrumplot}
\end{figure}

In Fig. \ref{sspectrumplot} we produce a plot of $\Delta^2 _{\mathcal{R}}$ for a range of low multipoles $2 \lesssim l \lesssim 28$, assuming that inflation is followed by instantaneous reheating. Different plots of the scalar power spectrum correspond to different values of an observationally relevant parameter $\mathfrak{R}$, which is  the dimensionless ratio of the fluid to inflaton energy densities when the dipole scale $l = 2$ exists the horizon. Fixing the value of $m$, we change $\mathfrak{R}$ by changing
the initial values of the inflaton field.

A number remarks are in order here:
\begin{itemize}

\item In Sec. \ref{vacuum} we argued that modes with wavenumbers $k
  \lesssim k_{\text{min}}$
have never been in an adiabatic regime since the Planck era.
This means that the dipole scale mode is not initially in an adiabatic
regime for $\mathfrak{R} > 1$. For the case of $\mathfrak{R} = 10$,
one can show that multipoles $l \gtrsim 5$ are initially in an
adiabatic regime.

\item The coupling of inflaton to radiation produces two main features
  in the scalar power spectrum: large amplitude suppression followed
  by small oscillations for low multipoles. The oscillations die off
  rapidly as $k$ increases and the scalar power spectrum quickly converges to the
  standard single field result. The period of oscillations, which is
  larger for larger values of $\mathfrak{R}$, can extend over a few
  multipoles. Nonetheless, the amplitude of oscillations does not
  exceed  a few percent.

\item Observational data for $l \lesssim 5$ are not reliable due to a
  high value of cosmic variance. For values of $\mathfrak{R} \gtrsim
  1$, one can expect deviations from the standard single field result
  for $l \gtrsim 5$, as indicated in Fig. \ref{sspectrumplot} for
  $\mathfrak{R} =10$. Even though the small oscillations produced in
  this model do not appear to be detectable, the
  sharp suppression at the end tail of the power spectrum could very
  well be. Nonetheless, this suppression is only slightly outside
  of the range of multipoles with large cosmic variance. Also one
  should bear in mind that these modes are barely adiabatic at the
  initial time.

\item Oscillatory features for low multipoles have previously been discovered in  a variety of models, including  trans-Planckian models \cite{easter1,danielsson1,martin1,bozza1}, axion monodromy models with additional instanton induced oscillatory features in the potential \cite{flauger1}, models with brief non slow-roll phases \cite{starobinsky1,hunt1,chung1,adams1,contaldi},
models of cascade inflation \cite{ashoor2,ashoor3}, and models with a sudden change in the speed of sound \cite{achucarro1,park,nakashima}. In our case, the oscillations are produced while the adiabaticity for all modes within the observable range as well as the slow-roll nature of the background dynamics  are preserved.

\item Sharp suppressions of the scalar power spectrum for small multipoles have been previously suggested in cutoff models \cite{sinha},
models of cascade inflation \cite{ashoor2, ashoor3}, as well as models with an initial non-slow-roll dynamics \cite{contaldi}.
Suppressions in the former class of models extend over a larger range of multipoles than in our case. However, the ones
observed in the latter class of models are quite similar to ours.

\item The features observed in this particular case study are in a good agreement with the
ones seen in the previous analyses of this model. In particular, both the oscillatory
features as well as the sharp suppressions pertaining to the low multipole scalar perturbation modes
were noticed in \cite{hirai} and \cite{kinney}. The large suppressions were also noticed in \cite{wang}.

\item Finally we comment on the effects of a non-trivial reheating
  scenario on the aforementioned results. A non-trivial reheating
  scenario, such as the canonical reheating mentioned in
  Sec. \ref{reheating}, results in a smaller ratio
$a_i/a_0$ of the initial scale factor and the scale factor today. Therefore,
the dipole momentum scale today would correspond to a larger momentum
scale when the fluid energy density is Planckian. Larger values of $k$
exit the horizon at later times, when the ratio of fluid energy
density over the inflaton energy density is smaller.  Hence lowering the temperature of reheating in
non-trivial reheating models correspond to having smaller deviations from the standard power spectrum at low multipoles.

\end{itemize}

\section{Power spectrum of gravitational waves} \label{tensorpower}

 In this section we compute the power spectrum of the stochastic
 background of gravitational waves produced in this model.
The quantization procedure for the gravitational waves in this model is quite standard and can be found in many references \cite{mukhanov,weinberg,baumann, dodelson}.

The power spectrum of the gravitational waves is defined as
\be \label{tpower}
 \langle \mathcal{J} |\hat{h}_{ij}(\bs{k} ,\eta_{\text{out}}) \hat{h}^{ij} (\bs{k'},\eta_{\text{out}}) | \mathcal{J} \rangle \equiv  \delta^{3}(\bs{k}+\bs{k'})\mathcal{P}_{h} (k),
\ee
where $\eta_{\text{out}}$ is the conformal time when $k = \mathcal{H}$, and $|\mathcal{J}\rangle$ is the Bunch Davies initial state. Here we have \cite{weinberg}
\be
\mathcal{P}_{h}(k) = 4 \frac{|\bar{h}(k, \eta_{\text{out}})|^2}{a^2 (\eta_{\text{out}})}.
\ee

Just as in the case of scalar power spectrum, it is convenient to define a dimensionless tensor power spectrum $\Delta^2 _{h}$ using
\be
\label{tdimp}
\Delta^2 _{h} \equiv \frac{k^3}{2 \pi^2} \mathcal{P}_{h}.
\ee
In the standard inflationary picture, due to the near scale invariance of the spectrum of tensorial fluctuations, one normally expresses $\Delta^2 _h$ using a simple ansatz
\be
\Delta^2_h = A_t \Big(\frac{k}{k_{*}} \Big) ^{n_t},
\ee
where
\be
A_t = \frac{2}{\pi^2} H^2
\ee
 is the value of the gravitational waves power spectrum evaluated at a pivot scale $k_{*}$, and
\be
n_t = -2 \varepsilon +\text{terms higher order in slow-roll},
\ee
is a measure of the scale dependence of the tensor power spectrum \cite{baumann}.

\subsection{Case study} \label{numerictensor}

\begin{figure}[h]
\centering
\includegraphics[width=0.9 \textwidth]{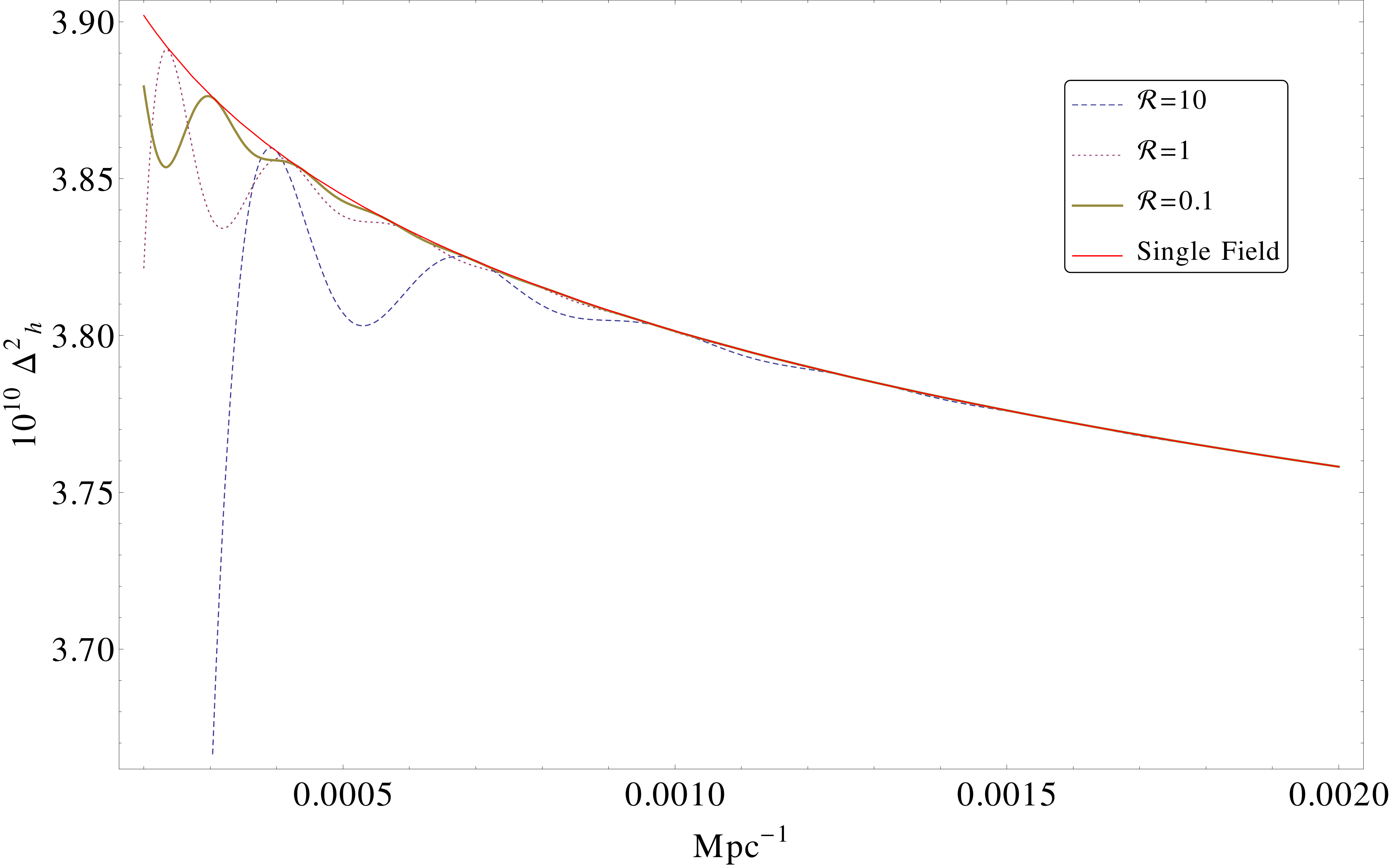}
\caption{The tensor power spectrum $\Delta^2 _h$ versus $k$ for a
  model with radiation coupled to an inflaton with the potential $V(\varphi) = 1/2  \ m^2 \  \varphi^2$ for $2 \lesssim l \lesssim 28$ assuming instantaneous reheating. The initial value of the inflaton field is slightly changed to produce different values of $\mathfrak{R}$.}
\label{tspectrumplot}
\end{figure}

We numerically compute the tensor power spectrum  from
Eq. \eqref{tdimp} for the same model studied in
Sec. \ref{numericscalar}.
We choose the Hubble parameter \cite{mcallister}
\be
H = 3 \times \sqrt{\frac{r}{0.1}} \times 10^{-5} \approx 4.24 \times 10^{-5}
\ee
for $l \approx 80$, which corresponds to a scalar to tensor ratio of
$r = 0.2$.
Using this value of the Hubble parameter, we calculate the amplitude $A_t$ of primordial gravitational waves  to be
\be
A_t  = \frac{2}{\pi^2} H^2 \approx 3.64 \times 10^{-10},
\ee
near $l \approx 80$. As in Sec. \ref{numericscalar}, we select values of the inflaton mass $m$  and the initial value of
the inflaton field that give both an $A_t$ close to the measured value near $l =80$ and $n_{t} \approx -2 \varepsilon$, as required from the theoretical considerations. Fixing the value of $m$, we  vary  the initial value of the inflaton field to produce various values of $\mathfrak{R}$. Fig. \ref{tspectrumplot} presents our result for different values of $\mathfrak{R}$.

We now make a number of remarks with regard to this result:
\begin{itemize}

\item For tensor perturbations, one can show by explicit calculation that the dipole scale is initially in an adiabatic regime  even
for $\mathfrak{R} = 10$. Recall that for $\mathfrak{R}>1$, the dipole scale is smaller than $k_{\text{min}}$.

\item The tensor power spectrum exhibits the same qualitative features as the scalar power spectrum; namely small oscillations
and sharp suppression for order unity multipoles.
The amplitude of oscillations however seem somewhat larger in this case, though not large enough
to merit a phenomenological significance.
Both of these features become insignificant for $l \gtrsim 5$.

\item  Similar oscillatory features were noticed in a study of trans-Planckian models in
\cite{ashoor}.

\item The observed features are in a good qualitative agreement with the ones found in an earlier study of this
model in \cite{hirai}.

\item As in the case of scalar spectrum, having a non-trivial model of reheating corresponds to less deviations from the standard result  in the observable window of scales.

\end{itemize}

\section{Concluding Remarks}

We recently learned that the mixed radiation-inflaton cosmological model studied in this paper had been previously studied with regard to
the issue of the robustness of inflationary predictions first in \cite{hirai} and subsequently
in \cite{kinney} and \cite{wang}. In \cite{hirai}, the authors computed the primordial scalar and tensor power
spectra, as well as the CMB angular TT and TE power spectra. They showed that both the primordial power
spectra and the CMB angular power spectra are suppressed at low multipole moments. They also found  both the scalar
and tensor primordial power spectra to have oscillatory features at low multipoles. A similar analysis
was subsequently performed in \cite{kinney}. There the authors computed  the scalar power spectrum as well as the CMB
angular TT and TE power spectra. Their analysis confirmed the conclusions of the previous analysis performed in \cite{hirai}.  Additionally, in
\cite{wang} the primordial power spectrum of a test scalar field in a fixed radiation-deSitter background
was computed, using which the CMB angular TT power spectrum was approximated. Both the
primordial and the CMB power spectra were found to be suppressed at low multipoles, though no oscillatory features
were noted.

In this work, we expand the previous analyses by a more detailed study of the background dynamics,
as well as a complete treatment of the scalar perturbations. We note here that in all of the
previous studies the scalar perturbation due to the radiation fluid was not properly included.
As such, the possibility of the generation of the entropic perturbations was not addressed.
In this work, by virtue of the scalar perturbation action derived in Sec. \ref{perturb}, we were able to identify
the exact form of the linear order couplings between the radiation and inflaton perturbation modes.
The study of these coupling functions together with the study of the background transition epochs
were instrumental in proving several key properties
of the perturbations. These include the arguments provided in Sec. \ref{iso}
for the suppression of the entropy perturbation modes.
Nonetheless, our analysis confirms the validity of the major conclusions previously drawn
 in \cite{hirai,kinney,wang}. In particular,
the results presented in Sec. \ref{numericscalar} for a
case study of the scalar perturbations is in a very good agreement
with the conclusions of \cite{hirai,kinney}, and to a lesser degree with those of  \cite{wang}.
Also, our results for  a case  study of the tensor perturbations presented in Sec. \ref{numerictensor}
are in good agreement with the ones found in \cite{hirai}.

Finally, we emphasize that our choice of a radiation fluid for the pre-inflationary phase is motivated by the fact that in this model the perturbations at sufficiently early times are adiabatic and so a natural choice of initial vacuum state exists.
However, just as with standard inflation, initial {\it non-vacuum states} may be physically relevant for a variety of reasons,
including the effect of trans-Planckian physics or the effects
of a phase that preceeds the adiabatic radiation-inflaton phase.
For standard inflation, one typically parametrizes trans-Planckian effects
on inflationary dynamics by  deviations from the initial Bunch-Davies vacuum
state\footnote{An initial thermal state could be a well motivated choice.
See \cite{star3,vilenkin1,vilenkin2}
where this choice of initial state is considered.}
 [see e.g. \cite{brand,greene} for some of the earliest approaches to this
problem]. While this is an interesting issue, we have not addressed it here as
it is beyond the scope of this paper.

\section{Acknowledgments}

This research was supported in part by
NSF grants PHY-1068541 and PHY-1404105 and by NASA grants
NNX11AI95G and NNX14AH53G.

\appendix

\section{Review of stress-energy tensor perturbations} \label{ptensor}

In this section we review how a general stress-energy tensor for a fluid can be linearly perturbed. For a perfect fluid, the stress-energy tensor can be written as
\be \label{stensor1}
T _{ab} = p g_{ab}+ (\rho+p )u_a u_b,
\ee
where $p$, $\rho$, and $u^a$ are the pressure, energy density, and the four-velocity of the fluid. An everywhere isotropic Universe  requires the components of this tensor to only depend on time.  However, allowing small spatial dependence for the fluid functions,
\bea
&& p(\eta,\bs{x}) \equiv p^{0} (\eta) + \delta p (\eta, \bs{x}), \hspace{1cm} \rho(\eta, \bs{x}) \equiv \rho^0 (\eta) + \delta \rho(\eta, \bs{x}), \nonumber\\
&& \hspace{2cm} u^{a} (\eta, \bs{x}) \equiv u_0 ^{a} (\eta) + \delta u^{a} (\eta, \bs{x}),
\eea
one is left with a slightly inhomogeneous and anisotropic Universe described by the linearly perturbed Friedmann-Robertson-Walker metric \cite{weinberg}
\be
ds^2 = - a(\eta)^2 \Big[ (1+ 2 \phi) d\eta^2- 2 B_i dx^i d\eta - [(1-2 \psi) \delta_{ij} + 2 E_{ij} ] dx^i dx^j \Big],
\ee
where $\phi$, $B_i$, $\psi$, and $E_{ij}$ are perturbation functions all being small compared to unity.

 We then define the background fluid four-velocity vector $u^a _0$ to be that of the frame in which the three-momentum density vanishes. Recalling that $g_{ab} u^a u^b = -1$ to all orders in perturbation, we define
\be
\vec{u}_0 \equiv   \bigg \langle \frac{1}{a}, 0,0,0 \bigg \rangle, \hspace{1cm} \vec{\delta u}  \equiv \bigg \langle -\frac{\phi}{a}, \frac{v^{i}-B^{i}}{a} \bigg \rangle,
\ee
where $(v^{i}-B^{i}) /a$ is the fluid three-velocity vector to leading non-vanishing order for some three-vector $\bs{v}$.

With these definitions, the stress-energy tensor \eqref{stensor1} becomes (to linear order in perturbations)
\bea
&& T_{00} = a^2 [ \rho^0 + \delta \rho + 2 \rho^0 \phi], \nonumber\\
&& T_{i0}= a^2 \big[ p^0 (B_i-v_i)- \rho^0 v_i \big], \nonumber\\
&& T_{ij} = a^2 \big[ p^0 \delta_{ij} + \delta p + 2  p^0  (E_{ij} - \psi) \big].
\eea
Finally we define a useful additive function known as the three-momentum density perturbations $\delta q^{i}$ using
\be
\delta q^{i} \equiv (p^0 + \rho^0) a v^i.
\ee
This quantity is proportional to the time-space component of the stress-energy tensor,
\be
T_{i} ^{0} = \frac{1}{a}\delta q_i.
\ee
For scalar perturbations, one can define a three-velocity potential as $\delta u_{,i} \equiv a v_i$. This in turn gives rise to the three-momentum density potential $\delta q$ defined as
\be
\delta q_{,i} \equiv (p^0 + \rho^0) \delta u _{,i}.
\ee

\section{Review of gauge transformations and gauge invariant functions}

In this section we will review the basics of gauge transformation in cosmological perturbation theory. We will then provide the gauge transformation relations for the perturbation functions used in this paper. For an excellent discussion of this concept see \cite{mukhanov} or a more recent review article \cite{baumann}.

Gauge transformation in cosmological perturbation theory is concerned with how  perturbation functions change under changes in  coordinate systems defined on some spacetime manifold. Given a set of coordinate system $(\eta, x^{i})$, we define an infinitesimal shift in coordinates as
\be \label{coordinatetrans}
\bar{\eta} \equiv \eta+ \xi^{0}(\eta,x^{i}) \hspace{1cm} \bar{x}^{i} \equiv x^{i} + \xi^{i} (\eta, x^{i}),
\ee
for some smooth functions $\xi^0$ and $\xi^{i}$. For scalar perturbations $\xi^{i} \equiv \partial^{i} \xi$, for some smooth function $\xi$.  The line elements defined in Eq. \eqref{pertmetr} are geometrical invariants, independent of a choice of coordinate system. We should therefore have
\be \label{metg}
ds^2 = g_{ab} (\eta,x^{i})  dx^{a} dx^{b} = \bar{g} _{ab}(\bar{\eta},\bar{x}^i)  d\bar{x}^{ a} d\bar{x}^{ b},
\ee
where $ \bar{g}_{ab} (\bar{\eta},\bar{x}^i) $ and $d\bar{x}^{ a} $ are the metric and differential 1-form defined using a new coordinate system $(\bar{\eta},\bar{x}^i)$. We can use Eq. \eqref{metg} to derive the changes in scalar metric perturbation functions,
\bea
&& \bar{\phi} = \phi - \mathcal{H} \xi^0 - \xi^{\prime 0 },\nonumber\\
&& \bar{\psi} = \psi + \mathcal{H} \xi^0 , \nonumber\\
&& \bar{B} = B+ \xi^0 - \xi', \nonumber\\
&& \bar{E} = E- \xi. \nonumber\\
\eea
Tensorial metric perturbations $h_{ij}$ are defined to be traceless and transverse, rendering them gauge invariant. Imposing the traceless-transverse condition ensures that no piece of $h_{ij}$ transforms as a scalar or a vector under the coordinate transformation \eqref{coordinatetrans}.

In addition to the metric perturbation functions, scalars defined on spacetime also undergo transformations upon changing coordinate systems. For a scalar function $\alpha(\eta, x^i)$, we can define its perturbation as
\be
\delta \alpha (\eta, x^i) \equiv \alpha(\eta,x^i) - \alpha_0 (\eta),
\ee
where $\alpha_0(\eta)$ is the background value of $\alpha(\eta,x^i)$. Keeping $\alpha(\eta,x^i)$ invariant under the coordinate transformation, we get
\be
\bar{\delta \alpha} = \delta \alpha - \alpha_0 ' \xi^{0}.
\ee
The scalar functions important in this paper are the fluid three-velocity potential $\theta$, the inflaton perturbation $\delta \varphi$, the three-momentum density potential $\delta q$ for the fluid and inflaton, and the energy density and pressure of both species. These scalar functions transform under an arbitrary coordinate transformation as
\bea
&& \bar{\theta}= \theta- \sqrt{2} z \mathcal{H} \xi^{0}, \nonumber\\
&& \bar{\delta \varphi} = \delta \varphi - \varphi' \xi^0,\nonumber\\
&&\bar{\delta q} = \delta q  - a (\rho^0+ p^0 ) \xi^{0}, \nonumber\\
&&\bar{ \delta \rho} = \delta \rho - \rho^{\prime 0} \xi^{0}, \nonumber\\
&& \bar{\delta p} = \delta p - p^{\prime 0} \xi^{0}.
\eea

Using the gauge transformation relations derived above for the metric perturbation functions and scalars, one can form gauge-invariant functions that are linear in these perturbations. The ones used in this paper are the mukhanov variables $\nu$ and $\pi$ for the fluid and inflaton,
\be
\nu \equiv \frac{1}{\sqrt{2}} (\theta - 2 z \psi) , \hspace{1cm} \pi \equiv a \delta \varphi + \zeta \psi,
\ee
the comoving curvature perturbation $\mathcal{R}$,
\be
\mathcal{R} \equiv \psi - \mathcal{H} \frac{\delta q}{a (\rho^{0} + p^{0} )},
\ee
and the entropy perturbation function $\mathcal{S}$
\be
\mathcal{S} \equiv \mathcal{H} \bigg( \frac{\delta \rho}{\rho^{\prime 0}} - \frac{\delta p}{p^{\prime 0}} \bigg).
\ee

\section{Quantization scheme for the scalar perturbations} \label{app2}

In this section we show how to properly quantize the scalar perturbations of our fluid-inflaton model. Our starting point is the scalar perturbation action
\begin{equation}\label{actmat}
\delta_2 S = \int \ d\eta \ \int d^3x \ \frac{1}{2} \Big[\bs{\Pi}^{\prime \text{T}} \bs{\Pi}' - \nabla_i \bs{\Pi}^{\text{T}} \bs{c}^2 _{s} \nabla^{i} \bs{\Pi} + \bs{\Pi}^{\text{T}} \bs{\Gamma} \bs{\Pi}  + \bs{\Pi}^{\prime \text{T}} \bs{\lambda} \bs{\Pi} \Big],
\end{equation}
where  the matrices are defined as
\begin{eqnarray} \label{matrices}
&& \bs{\Pi} \equiv \begin{pmatrix}
                                        \pi \\
                                        \nu
                                        \end{pmatrix},
\hspace{1 cm}   \bs{c}^2 _{s} \equiv \begin{pmatrix}
                         1 & 0 \\
                        0 & c_s ^2
                        \end{pmatrix},
\hspace{1cm}  \bs{\lambda} \equiv \begin{pmatrix}
                                                        0 & \sqrt{2} \frac{B}{\zeta} \\
                                                        - \sqrt{2} \frac{B}{\zeta}&0 \\
                                                        \end{pmatrix}  , \nonumber\\
&&  \bs{\Gamma} \equiv \begin{pmatrix}
                                        \frac{\zeta''}{\zeta} + \frac{2}{\zeta} \big[ 4 z^2 A - (z B)' \big] & 4 \sqrt{2} A \frac{z}{\zeta}-\frac{\sqrt{2}}{2} B \frac{\zeta'}{\zeta^2}-\frac{\sqrt{2}}{2} \frac{B'}{\zeta} \\
                                        4 \sqrt{2} A \frac{z}{\zeta}-\frac{\sqrt{2}}{2} B \frac{\zeta'}{\zeta^2}-\frac{\sqrt{2}}{2} \frac{B'}{\zeta} & \frac{z''}{z}+4A \\
                                        \end{pmatrix}.
\end{eqnarray}
We have suppressed dependences on $\eta$ for all background functions above. The Euler-Lagrange equations of motion for this action are
\be \label{lageqm}
\bs{\Pi}''- \big[ \bs{c}_s ^2 \nabla^2 + \bs{\Gamma}-\frac{1}{2}\bs{\lambda}' \big] \bs{\Pi}+ \bs{\lambda} \bs{\Pi}'=0.
\ee

Next we compute the Hamiltonian for this system. First note that the canonical momentum  $\mathbf{P}$ in this model is
\begin{equation}
\mathbf{P} \equiv \frac{\delta \text{L}_{\text{scalar}}}{\delta \bs{\Pi}^{\prime \text{T}}} = \bs{\Pi}' + \frac{1}{2} \bs{\lambda} \bs{\Pi},
\end{equation}
where $\text{L}$ is the Lagrangian  appearing in  the action \eqref{actmat}. Employing the Legendre transformation, the Hamiltonian  becomes
\bea \label{hamiltonian}
&& \text{H}_{\text{scalar}} = \int \ dx^3   \ \big[ \mathbf{P}^{\text{T}} \bs{\Pi}' - \mathcal{L} \big] \nonumber\\
&& = \int \ dx^3 \ \frac{1}{2} \bigg[\mathbf{P}^{\text{T}}\mathbf{P}+ \nabla_i \bs{\Pi}^{\text{T}} \bs{c}^2_{s} \nabla^{i} \bs{\Pi} - \mathbf{P}^{\text{T}} \bs{\lambda} \bs{\Pi}+ \bs{\Pi}^{\text{T}} \big( \frac{1}{4} \bs{\lambda}^{\text{T}} \bs{\lambda} - \bs{\Gamma} \big) \bs{\Pi} \bigg],
\eea
which in  Fourier space reduces to
\bea \label{hamilk}
\text{H}_{\text{scalar}} = \int \ dk^3 \ \frac{1}{2} \big[\mathbf{P}^{\text{T}}_{\bs{k}}  \mathbf{P}_{-\bs{k}}+ \bs{\Pi}^{\text{T}} _{\bs{k}} \bs{\Sigma} \bs{\Pi}_{-\bs{k}} - \mathbf{P}^{\text{T}} _{\bs{k}} \bs{\lambda} \bs{\Pi}_{-\bs{k}} \big].
\eea
Here we define
\begin{equation}
\bs{\Sigma} \equiv \bs{c}^2 _s k^2 -\bs{\Gamma}+ \frac{1}{4} \bs{\lambda}^{\text{T}} \bs{\lambda}
\end{equation}
( Our convention for Fourier transforming functions is $X(\bs{x},\eta)  = \int \ dk^3/(2 \pi)^{3/2}  \  X(\bs{k},\eta) e^{i \bs{k}. \bs{x}}$). The Hamilton's equations of motion for each mode are
\bea \label{hamileqm}
&&\mathbf{P}'_{\bs{k}} = -\frac{\delta \text{H}}{\delta \bs{\Pi}^{\text{T}}_{-\bs{k}}}= - \bs{\Sigma} \bs{\Pi}_{\bs{k}}- \frac{1}{2}  \bs{\lambda} \mathbf{P}  _{\bs{k}}, \nonumber\\
&& \bs{\Pi}' _{\bs{k}} = \frac{\delta \text{H}}{\delta \mathbf{P}^{\text{T}}_{-\bs{k}}}= \mathbf{P}_{\bs{k}}-\frac{1}{2} \bs{\lambda} \bs{\Pi}_{\bs{k}}.
\eea

The Hamiltonian \eqref{hamilk} is the starting point for quantization. After promoting $\mathbf{P}_{\bs{k}}$ and $\bs{\Pi}_{\bs{k}}$ to quantum operators, we impose the following equal-time quantization relations,
\bea
&& \big[\hat{\Pi}_{i\bs{k}}, \hat{\Pi}_{j\bs{k}'} \big]=0 , \hspace{1 cm} \big[\hat{\text{P}}_{i \bs{k}}, \hat{\text{P}}_{j\bs{k}'} \big]=0 , \hspace{1 cm}  \big[\hat{\Pi}_{i \bs{k}}, \hat{\text{P}}_{j\bs{k}'} \big]= i \delta_{ij} \delta^{3} (\bs{k}+\bs{k}').
\eea

Quantum operators such as $\hat{\mathbf{P}}$ and $\hat{\bs{\Pi}}$ are often conveniently expanded in terms of the creation and annihilation operators and the mode functions that are solutions to the equations of motion \eqref{lageqm} or \eqref{hamileqm}. Note that the space of solutions of \eqref{lageqm} is four-dimensional. Therefore, we can pick any two linearly independent solutions $\bs{\Pi}_{A}$ and $\bs{\Pi}_{B}$ together with their complex conjugates to span the space of (Lagrangian) solutions. This leads to the expansion of the quantum operator $\hat{\bs{\Pi}}$  as
\be\label{quantexp}
\hat{\bs{\Pi}}_{\bs{k}}(\eta) = \bs{\Pi}_{A}(k,\eta) \hat{a}_{A,\bs{k}} + \bs{\Pi}^{*} _{A} (k, \eta) \hat{a}^{\dagger} _{A, -\bs{k}},
\ee
where $\hat{a}$ and $\hat{a}^{\dagger}$ are the creation and annihilation operators, and summation over the index $A$ is assumed.

It remains to give a proper definition for the creation and annihilation operators $\hat{a}$ and $\hat{a}^{\dagger}$. To do this, we define the inner product on the space of solutions of \eqref{lageqm} as
\be
\langle \mathbf{U}, \mathbf{V} \rangle \equiv i \big[ \mathbf{U}^{+} \mathbf{V}'-\mathbf{U}^{\prime +}\mathbf{V}+ \mathbf{U}^{+} \bs{\lambda} \mathbf{V} \big],
\ee
where $+$ denotes the Hermitian conjugate of a matrix. Using the equations of motion \eqref{lageqm} one can show this inner product to be conserved, i.e. $\partial _{\eta} \langle \bs{U}, \bs{V} \rangle=0$, for all solutions, making it an appropriate definition of inner product on the space of Lagrangian solutions. We then require the solutions $\bs{\Pi}_{A} (k, \eta)$ appearing in \eqref{quantexp} to be normalized in the sense of
\be \label{norm}
\big \langle \bs{\Pi}_{A}(k, \eta) , \bs{\Pi}_{B} (k, \eta) \big \rangle = \delta_{AB}.
\ee
Conventionally we pick
\begin{equation} \label{initial}
\bs{\Pi}_1 (k,\eta) = \begin{pmatrix}
                        \frac{e^{-i k \eta}}{\sqrt{2k}}\\
                        0
                        \end{pmatrix}, \hspace{ 1cm} \bs{\Pi}_2 (k, \eta) = \begin{pmatrix}
                        0 \\
\frac{e^{-i c_s k  \eta}}{\sqrt{2 c_s k}}
                        \end{pmatrix}.
\end{equation}
These two solutions are linearly independent, normalized in the sense of \eqref{norm},  and together with their complex conjugates form a basis for the space of solutions of \eqref{lageqm} at early times, when the fluid and  inflaton scalar perturbations are both decoupled and adiabatic. These solutions correspond to the fluid and inflaton perturbation modes being in some vacuum state at early times \cite{mukhanov}.

Finally we define the creation and annihilation operators as
\be
\hat{a}_{A,\bs{k}} \equiv \big \langle \bs{\Pi}_{A} (k,\eta), \hat{\bs{\Pi}}_{\bs{k}}(\eta) \big  \rangle, \hspace{1cm} \hat{a}^{\dagger} _{A,\bs{k}} \equiv - \big \langle  \bs{\Pi}^{*} _{A}(k,\eta) ,\hat{\bs{\Pi}}_{-\bs{k}}(\eta) \big \rangle.
\ee
It is easy to check  that
\be
[ \hat{a}_{A,\bs{k}}, \hat{a}^{\dagger}_{B,\bs{k}'} ] = \delta_{AB} \delta^{3} (\bs{k}-\bs{k}') , \hspace{1cm} [ \hat{a}_{A,\bs{k}}, \hat{a}_{B,\bs{k}'} ] = [ \hat{a}^{\dagger} _{A,\bs{k}}, \hat{a}^{\dagger}_{B,\bs{k}'} ] =0 .
\ee
The operators $\hat{a}_{A\bs{k}}$ annihilate the ground states $|\Omega_{A} \rangle$ defined as
\be \label{groundstate}
 |\Omega_{A} \rangle  \equiv \bigotimes_{\bs{k}} |0_{A\bs{k}} \rangle,
\ee
where $ |0_{A\bs{k}} \rangle$ denotes the vacuum state corresponding to the mode with momentum $\bs{k}$. Existence of a natural choice of vacuum at early times is the consequence of the adiabaticity of fluid and inflaton perturbation modes discussed in Sec. \ref{properties}. Indeed the vacuum $|\Omega_{1(2)}\rangle$ is the analog of the Bunch-Davies  state for the inflaton (fluid) perturbations, corresponding to mode functions with initial conditions \eqref{initial}. The vacuum state for the scalar perturbations is then defined as
\be
|\mathcal{I}\rangle \equiv \bigotimes_{A} |\Omega_{A}\rangle.
\ee

\bibliographystyle{JHEP}
\bibliography{Reference}

\end{document}